# Numerical Fitting-based Likelihood Calculation to Speed up the Particle Filter

Tiancheng Li, Shudong Sun, Juan M. Corchado, Tariq P. Sattar and Shubin Si

T. Li is with the with the research group of Bioinformatics, Intelligent Systems and Educational Technology (BISITE), Faculty of Science, University of Salamanca, 37008 Salamanca, Spain. T. Li is also with the School of Mechanical Engineering, Northwestern Polytechnical University, Xi'an, 710072, China. T. Li will correspond to this contribution; e-mail: tiancheng.li1985@gmail.com; webpage: https://sites.google.com/site/tianchengli85/).

S. Sun is with the School of Mechanical Engineering, Northwestern Polytechnical University, Xi'an, 710072, China (e-mail: sdsun@nwpu.edu.cn).

J. M. Corchado is with the BISITE group, Faculty of Science, University of Salamanca, 37008 Salamanca, Spain. (e-mail: corchado@usal.es).

T. P. Sattar is with the Centre for Automated and Robotics NDT, London South Bank University, SE1 0AA, UK (e-mail: tpsattar@lsbu.ac.uk)

S. Si is with the School of Mechanical Engineering, Northwestern Polytechnical University, Xi'an, 710072, China (e-mail: sisb@nwpu.edu.cn).

**Abstract**

The likelihood calculation of a vast number of particles is the computational bottleneck for the particle filter in applications where the observation information is rich. For fast computing the likelihood of particles, a numerical fitting approach is proposed to construct the Likelihood Probability Density Function (Li-PDF) by using a comparably small number of so-called fulcrums. The likelihood of particles is thereby analytically inferred, explicitly or implicitly, based on the Li-PDF instead of directly computed by utilizing the observation, which can significantly reduce the computation and enables real time filtering. The proposed approach guarantees the estimation quality when an appropriate fitting function and properly distributed fulcrums are used. The details for construction of the fitting function and fulcrums are addressed respectively in detail. In particular, to deal with multivariate fitting, the nonparametric kernel density estimator is presented which is flexible and convenient for implicit Li-PDF implementation. Simulation comparison with a variety of existing approaches on a benchmark 1-dimensional model and multi-dimensional robot localization and visual tracking demonstrate the validity of our approach.

# 1 INTRODUCTION

This paper concerns the design of a high-speed particle filter (PF) for a variety of nonlinear state estimation problems such as localization, positioning and tracking. Computing efficiency is a critical requirement in industry but is particularly challenging for the application of the particle filter. In brief, the nonlinear filtering recursively estimates the nonlinear sequence of posterior densities of the state given a sequence of observations, which can be written in the form of the discrete dynamic State Space Model (SSM)

$$\begin{cases} x_t = f_t(x_{t-1}, u_t) & \text{(state transition equation)} \\ y_t = h_t(x_t, v_t) & \text{(observation equation)} \end{cases} \quad (1)$$

where $t$ indicates discrete time, $x_t \in \mathbb{R}^{d_x}$ denotes the state, $y_t \in \mathbb{R}^{d_y}$ denotes the observation, $u_t$ and $v_t$ denote stochastic noise affecting the state transition equation $f_t: \mathbb{R}^{d_x} \times \mathbb{R}^{d_u} \to \mathbb{R}^{d_x}$, and the observation equation $h_t: \mathbb{R}^{d_x} \times \mathbb{R}^{d_v} \to \mathbb{R}^{d_y}$, respectively. Furthermore, let $x_{0:t} \triangleq (x_0, x_1, \ldots, x_t)$ and $y_{0:t} \triangleq (y_0, y_1, \ldots, y_t)$ be the history path of the signal and of the observation process respectively.

A standard and convenient solution to the SSM-based filtering problem is the *Recursive Bayesian estimation*, which is based on two assumptions as follows:

(A.1) The states follow a first-order Markov process

$$p(x_t | x_{0:t-1}) = p(x_t | x_{t-1}) \quad (2)$$

(A.2) The observations are independent of the given states

$$p(y_{0:t-1} | x_t) = \frac{p(x_t | y_{0:t-1}) p(y_{0:t-1})}{p(x_t)} \quad (3)$$

Using the *Bayes' rule*, the required marginal posterior density is found as

$$p(x_t | y_{0:t}) = \frac{p(y_t | x_t) p(x_t | y_{0:t-1})}{p(y_t | y_{0:t-1})} \quad (4)$$

where $p(y_t | x_t^{(i)})$ is the likelihood. This can be determined in two steps that form one iteration:

(Step.1) Prediction (Chapman-Kolmogorov equation)

$$p(x_t|y_{0:t-1}) = \int_{R^{nx}} p(x_{t-1}|y_{0:t-1})p(x_t|x_{t-1})dx_{t-1} \quad (5)$$

(Step.2) Updating or correction (Bayes' rule)

$$p(x_t|y_{0:t}) = \frac{p(x_t|y_{0:t-1})p(y_t|x_t)}{\int_{R^{nx}} p(x_t|y_{0:t-1})p(y_t|x_t)dx_t} \quad (6)$$

However, these two steps is only conceptual and the involved integration is generally incomputable. To solve the integration, we have to turn to suboptimal simulation-based methods, such as the Point-Mass (PM) filter [1, 2], the unscented filter [3] and Monte Carlo methods. The Monte Carlo method has become one of the standard and popular tools for sophisticated models, typically including Markov Chain Monte Carlo (MCMC) [12] and Sequential Monte Carlo (SMC) (often called PF; staged reviews of the state of art can be found in [4, 5, 6, 7, 8]), etc. Specifically in PF, the posterior density is represented by a set of particles with associated weights, and generally the computation complexity is directly proportional to the number of particles used as each particle needs at least to execute the state-prediction and weight-updating steps in parallel.

The PF has been widely used for nonlinear and non-Gaussian SSMs. However, it suffers from heavy computation due to a vast number of particles necessarily used, which is its primary disadvantage as compared to closed-form solutions like Kalman filters. In real-life applications, the weight-updating step is hardware-sensitive and is often more computationally intensive than the state-prediction step. This is particularly true for positioning [8, 9], visual tracking [10-11], robot localization [14-16] and Machine prognosis [17-18]. In these applications, the estimation accuracy and computing speed are both restricted heavily by the updating step. Based on this fact, this paper investigates speeding up the updating step so that to speed up PF without reducing the estimation quality.

We propose a numerical fitting approach to calculate the likelihood of particles that does not need to reduce the number of particles. Our approach is based on the understanding that the direct likelihood calculation based on observations is computationally more intensive as compared to the numerical fitting used.

Numerical fitting has proved to be a powerful and universal method for data prediction and has been used in a range of statistical applications where adequate analytical solutions may not exist. In the field of Kalman filtering, the well-known unscented transform technique is a proven type of statistical linear regression [3]. To our knowledge, this is the first attempt to employ the numerical fitting technique for likelihood calculation for PF. Both the ideas of inferring the likelihood of particles by others and employing numerical fitting to construct arbitrary PDF for particle filtering are new.

The remainder of this paper is organized as follows. A brief review of the state of the art in the development of fast processing PF is described in section 2. The conceptual framework and implementation details of the proposed approach are given in section 3. Simulations are presented in section 4 and the conclusion is given in section 5.

## 2 Brief review: Fast Particle Filtering

The primary obstacle for the application of the particle filter is from its computational inefficiency. One of the most straightforward and effective solutions to improve its speed is to minimize the required number of particles, see [19-20]. However, caution has to be exercised on reducing the number of particles as a smaller number of particles make it hard to approximate the underlying Probability Distribution Function (PDF) properly and to cope with the information imprecision. Several advanced forms of the particle filter have been proposed to work well with fewer samples, but it is seldom possible to get a win-win situation for general cases. For example, the cost reference PF [10] work with as few as ten particles for the 1-demensional estimation. In [11] the importance density in particles can be modified to interpret the posterior state by using pseudo-likelihoods to reduce the number of particles required. The box particle [13] occupies a small and controllable rectangular region in the state space, which reduces the computational complexity in high dimensional problems and is suitable for parallel processing.

A typical assumption underlying PF is that all samples can be updated whenever new sensor information arrives. Under real-life conditions, however, it is possible that the update may not be completed before the next sensor observation arrives. This can be the case for computationally complex sensor models or whenever the underlying posterior requires large sample sets, see [19]. To avoid the loss of observations when the rate of incoming sensor data is higher than the updating rate, a mixture of individual sample sets are used by distributing the samples among the observations within an update window. It is fair to say that the best way to avoid information loss is to improve the sampling speed so that observations obtained by sensors can be maximally employed.

Efforts have been made to increase the filtering speed by simplifying the updating step that creates the main computational burden. This can be achieved in two ways. One way is to reduce the number of required updating cycles. The other way is to reduce the computation of each updating cycle. In the first way, only observations that fall inside a specific scope around the particle have significant impact to the weight of the particle, while those outside of the scope have negligible impact and are therefore not taken into account [21]. This reduces the number of cycles of computationally intensive updating without reducing the number of particles used. In the second method, solutions are proposed to deal with observations for fast processing. For example, an adjustable observation model is proposed in [22] that can change between connected component analysis and k-means to obtain a balance between tracking precision and reduced runtime.

Real-time techniques such as parallel processing, dimension decomposition, multi-resolution processing, etc. provide possibilities for fast processing of PF. The essence of distributed PF is to distribute the algorithm among different computing agents for fast parallel computing. Given the fast development of computers, multicore platforms and general purpose graphics processing units are now available on almost every computer, and distributed computing becomes more popular and promising [24, 25]. One of the biggest challenges for developing parallel PF is the resampling operation that requires the joint processing of all particles and therefore prevents parallelization of PFs. To combat or to avoid this, various solutions for parallel resampling and parallel particle filtering have been proposed, see [23]. There is another type of PF which assume the underlying distribution is Gaussian and thereby resampling can be avoided [26].

The dimension- decomposition idea of Rao–Blackwellization (RB) [27] is to divide the state so that the

Kalman filter is used for the part of the state that is linear, and PF is used for the other part that is nonlinear, which inspires many similar developments to reduce the dimensions of the state space that needs to be processed. For example, the proposed method estimates orientation by using a particle filter, while the position and velocity is estimated by using KF [9]. Furthermore, to remove the linear limitation of Kalman filter, the Decentralized PF (DPF) [28] splits the filtering problem into two nested sub-problems and handles each individually using PFs. This differs from RB in the manner that two parts are all approximated by conditional PFs.

Furthermore, one may partition the state space into more subspaces and run separate PFs in each subspace [29]. A similar idea is implemented in [6] which represents each component as a single chain Bayesian network and uses PF to track each component for multi-component tracking. Similarly, the so-called partitioned sampling consists of dividing the state space into two or more partitions and sequentially applying the dynamics for each partition followed by an appropriate weighted resampling operation [30]. Splitting the state space is also an appealing and even necessary way to deal with high dimensionality [43]. Meanwhile, the time scale separation exhibited in [31] allows two simplifications of PF: 1) to use the averaging principle for the dimensional reduction of the dynamics for each particle during the prediction step and 2) to factorize the transition probability for the RB of the update step. The resulting PF is faster and has smaller variance than the original PF. On the other side, MCMC is more effective than PFs in high-dimensional spaces and therefore can be employed to benefit the PF [12] but it is not so suitable for online calculation. It is fair to say that high-dimensionality remains challenging for the application of PFs [32, 42, 43].

## 3 THE CORE IDEA: LI-PDF BASED WEIGHT UPDATING

### 3.1 The conceptual framework

The PF evaluates the posterior PDF by a set of particles $x_t^{(i)}$ with associated non-negative weight $w_t^{(i)}$ that employs the strong law of large numbers (SLLN), i.e.

$$p(x_t|y_{0:t}) \propto p(y_t|x_t) \sum_{i=1}^{N} w_{t-1}^{(i)} p(x_t|x_{t-1}^{(i)}, y_{0:t-1}) \quad (7)$$

where N is the number of particles, $t=0$ denotes the initialized particles set. The weight of particles is determined based on *Sequential Importance Sampling* (SIS)

$$w_t^{(i)} \propto w_{t-1}^{(i)} \frac{p(y_t|x_t^{(i)}) p(x_t^{(i)}|x_{t-1}^{(i)})}{q(x_t^{(i)}|x_{0:t-1}^{(i)}, y_{1:t})} \quad (8)$$

where $q(\cdot)$ is the proposal important density, also called evidence, $p(y_t|x_t^{(i)})$ is the likelihood of the particle $x_t^{(i)}$ given observation $y_t$ which is critical to the PF. There is a variety of ways to design efficient important function and sampling methods. For simplicity, the proposal density is often chosen as $p(x_t|x_{t-1})$ which minimizes the variance of the importance weights conditional upon the simulated trajectory $x_{0:t-1}^{(i)}$ and the observation $y_{1:t}$ [27]. Moreover, *resampling* may be applied to reset particles' weight [23] to be equal or approximately equal so that to combat weight degeneracy, i.e. the classical *Sampling Importance Resampling* (SIR), which is also referred to as *branching and interacting particle system* [33].

This study focuses on numerically fitting the *Probability Density Function* of the *Likelihood* (Li-PDF) and then use it to calculate the likelihood of each particle

$$p(y_t|x_t^{(i)}) = L_t(x_t^{(i)}) \quad (9)$$

where $L_t(\cdot)$ is the Li-PDF at time $t$, which can be either explicit or implicit as defined later in this paper. The following two definitions are at the core of our approach.

**Definition 1**. The task of numerical fitting is to recover $y = f(x; C)$, where $C$ are the parameters that are to be determined by using a given data based on the belief that this data contains a slowly varying component, which captures the trend of, or the information about, $y$, and a varying component of comparatively small amplitude which is the error or noise in the data. There are two forms of numerical fitting: regression and interpolation, which are distinguished from one another based on whether the function works on the data (that is interpolation) or not (regression).

**Definition 2**. The given data points used for fitting in our approach are called fulcrums. Fulcrums and particles have the same characteristics, see section 3.2.

Simply stated, the likelihood of the particles is obtained by fitting the likelihoods of the fulcrums in the Li-PDF approach, as depicted in Fig. 1. In Fig. 1, the horizontal axis and the vertical axis represent the state and likelihood respectively. Circles represent the particles while boxes represent the specific fulcrums and their likelihoods (the red dotted lines) are known. Firstly, the likelihoods of the fulcrums are fitted with their state to get the Li-PDF, as represented by the red curve. This curve is then used to obtain the particles' likelihood i.e. the height of black lines. In this way, no matter how many particles there are, we can fit all of them to get their likelihoods instead of direct calculation based on the observation.

The present approach implicitly assume that the likelihood function is continuous and has a smooth distribution, which is the case in most applications. The framework of the explicit Li-PDF based particle filter is described in Algorithm 1. The details of the algorithm are given in the following subsections.

Remark 1. Our approach does not have to be based on the SIR filter but it can be also combined with advanced PFs such as Auxiliary PF [34] and Gaussian PF [26]. Also, the Li-PDF approach follows no single set form but may instead have multiple implementation stages. Section 3.5 will show that the observation function (instead of the likelihood PDF) might be fitted firstly, after which only one more step of conversation is required to obtain the final likelihood PDF. Furthermore, the fitting of the observation function can be implemented in a batch manner for further speeding up with accuracy.

Remark 2. It has become a common idea to develop analytic technologies to improve random-sampling based approximation of PF, including the regularized PF (RPF) [35], kernel PF (KPF) [36], convolution PF [37] and feedback PF [38], but with very different implementations and purposes. In addition, Gaussian mixtures (GMs) are used [39] to represent the posterior PDF and the observation likelihood function. The continuous function in the form of either kernels [35, 36] or GMs [39] is propagated over time and is used to calculate the estimate. These PFs do not run faster than SIR PF if the same number of particles is used. In contrast, the likelihood function constructed online in our approach will not be propagated over time and no special assumption such as Gaussian distribution is required and a faster speed is achieved.

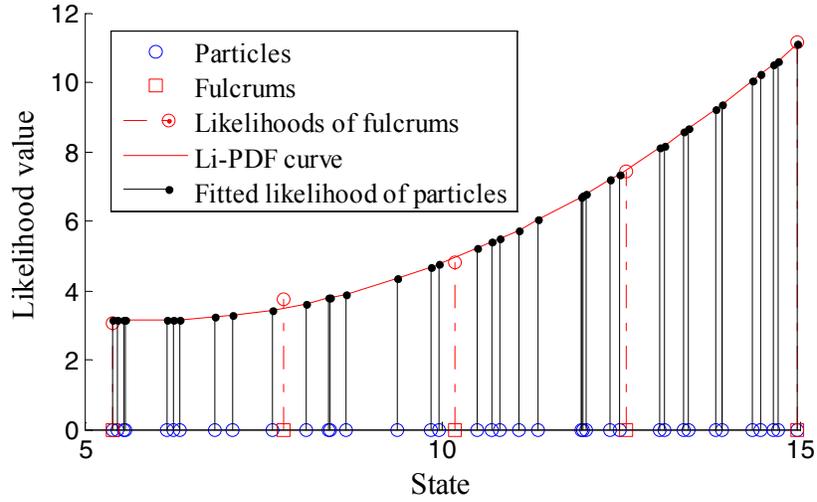

Fig.1 Schematic diagram of the Li-PDF approach for likelihood calculation

*Algorithm 1:* Explicit Li-PDF based PF (one iteration)

Input: $S_{t-1} = \{x_{t-1}^{(i)}, w_{t-1}^{(i)}\}_{i=1}^{N}$, Output: $S_t = \{x_t^{(i)}, w_t^{(i)}\}_{i=1}^{N}$

1. **Selective Resampling (**do if the variance of normalized weights is greater than a pre-specified threshold**)**

   Unbiased resampling that equally weight resampled particles e.g. systematic/residual resampling [23] is preferred

2. **Prediction**

For $i = 1 \rightarrow N$, sample from the proposal:

$$x_t^{(i)} \sim q\left(x_t^{(i)} \big| x_{t-1}^{(i)}, y_t\right)$$

3. **Li-PDF construction**

   **3.1** Construct $M$ fulcrums (see subsection 3.2): $(X_1, X_2, ... X_M)$

   **3.2** Calculate their likelihoods: $Y_m = p(y_t|x_t)$ of $X_m$

   **3.3** Numerically fit the likelihood of fulcrums with their states to get the Li-PDF $L_t(\cdot)$ (see subsections 3.3/3.4), which satisfies:

$$(Y_1, Y_2, \ldots Y_M) \cong L_t(X_1, X_2, \ldots X_M)$$

**4. Updating**

Update the weights by using their fitted value of $L_t(\cdot)$, i.e. for $i = 1 \to N$, update the weight:

$$w_t^{(i)} \propto w_{t-1}^{(i)} \frac{L_t\left(x_t^{(i)}\right) p\left(x_t^{(i)} \mid x_{t-1}^{(i)}\right)}{q\left(x_t^{(i)} \mid x_{0:t-1}^{(i)}, y_{1:t}\right)}$$

Normalize the weights: $w_t^{(i)} = w_t^{(i)} / \sum_{j}^{N} w_t^{(j)}$

*3.2 Non-negligible support fulcrums*

There are two methods to construct fulcrums: One method is to simply select some particles from the particle set (non-uniformly distributed) and the other is to create new data-points in the state space (uniformly distributed). According to the SLLN, more fulcrums are more likely to get a better fitting approximation, but at the higher cost of computation. The fulcrums should be distributed appropriately so that they have an adequate representation of all the particles with the fewest possible fulcrums. In our approach, a grid-based method is adopted to generate uniformly distributed fulcrums that cover the non-negligible region

By partitioning the state space of particles into rectangular cells, the fulcrums can be easily created by the centers of those cells. This method of approximating the probability density by rectangular delimited data-points is flexible and convenient for implementation (the data structure created is easy to handle in computers), which is subject to the independence between different dimensions. This data model has also been employed in the PM filter. Specifically, the anticipative boundary-based grid and the non-negligible support principle proposed in [1, 2] are readily suitable for our approach by a sensible albeit convenient conversion from the predictive PDF in the PM filter to the Li-PDF in the PF.

In contrast to interpolation, which predicts within the range of values in the dataset used for model fitting, prediction outside this range of the data is known as extrapolation. The further the extrapolation goes outside the data, the more likely it is for the model to fail due to differences between the fitting assumptions and the sample data or the true values. To avoid this, fulcrums are constructed in the state space $I_t$ to cover the

complete state-space of particles with a boundary margin $r$

$$\forall l \in [1, L]: I_t^{[l]} = \left[ \min_{\text{particles}} x_t^{[l]} - r^{[l]}, \max_{\text{particles}} x_t^{[l]} + r^{[l]} \right] \quad (10)$$

where $x_t^{[l]}$ is the state value in the $l$th coordinate, $L$ is the total dimensionality, $r^{[l]}$ is the $l$th dimension boundary margin which will be determined on the observation noise as in [1]

$$r^{[l]} = a\sqrt{Q_t^{[l]}} \quad (11)$$

where $Q_t$ is the observation noise covariance and $a$ is the parameter that determines the non-negligible level.

It remains to be shown how many fulcrums are required. One adaptive technique for setting the number of grid points [1] is that the number of data points $M_t^{[l]}$ satisfies

$$M_t^{[l]} \geq \frac{\|I_t^{[l]}\|}{2\gamma} \left(Q_t^{[l]}\right)^{-1/2} \times \max_{x_t \in \Omega_t} \left| \det \frac{\partial h(x_t)}{\partial x_t} \right| \quad (12)$$

where $\Omega_t$ is a significant support of the predictive PDF $p(x_t|y_{t-1})$, $h(x)$ is the observation function, $\gamma > 0$ is the second design parameter. In our approach, we adopt a much simpler calculation form as follows

$$M_t^{[l]} \geq \frac{\|I_t^{[l]}\|}{2\lambda} \left(Q_t^{[l]}\right)^{-1/2} \quad (13)$$

where $\lambda$ is a modified parameter to replace $\gamma$ with consideration of the observation equation. This can be determined offline according to our preference of the accuracy.

Since the fulcrum is the crossing point of the partitioning of each dimensionality, the total number of fulcrums is just the product of the partitioning number of each dimensionality

$$M_t = \prod_{l=1}^{L} M_t^{[l]} \quad (14)$$

If we delimit the fulcrums with a fixed interval $d^{[l]}$, i.e.

$$d^{[l]} = \frac{I_t^{[l]}}{M^{[l]} - 1} \quad (15)$$

Then, the fulcrums can be defined at the space crossing of the following coordinates:

$$\forall m \in \left[1, M_t^{[l]}\right]: x_{t,m}^{[l]} = \min_{\text{particles}} x_t^{[l]} - r^{[l]} + (m-1)d^{[l]} \quad (16)$$

There is no doubt that the larger the observation noise, the more fulcrums it requires. Choosing a sensible number of fulcrums with respect to the observation noise is important in our approach. For simplicity and fast online computation in multiple dimension situations, the following number $M^{[l]}$ is suggested in our application such that

$$\forall l \in [1, L]: M_t^{[l]} = p \text{ or } 1 \quad (17)$$

where $p$ is a specified value loosely satisfying (13), $M^{[l]} = 1$ means the insignificant $l$th dimensionality is not partitioned.

To note, fulcrums can be added into the particle set. This will not increase additional likelihood computation as the likelihood of the fulcrums has already been calculated. Obviously, the total number of particles that will affect the computation of other parts of the filter may be increased, unless a solution is found to remove some unwanted particles.

**Remark 3.** The primary limitation of the grid partitioning is due to the sensitivity to the dimensionality of the state space. To mitigate this limitation, partitioning the state space only in part/primary dimensions is highly recommended. For example, in the case of the state that consists of position, velocity, etc., the grid partitioning can be realized only in the position space (as shown in the robot localization in Section 4.2).

*3.3 Least squares numerical fitting*

Numerical fitting is accomplished in practice by selecting a linear or nonlinear function

$$y = f(x; c_1, c_2, ..., c_k) \quad (18)$$

that depends on certain parameters $c_1, c_2, ..., c_k$. It should be noted that the fitting data may not strictly work on the function, instead a fitting error generally exists, i.e.

$$y_m = f(x_m; c_1, c_2, ..., c_k) + e_m \quad (19)$$

where $y_m$ is the measured value of the dependent variable, $c_1, c_2, ..., c_k$ are the required parameters. In our approach, the given data $(x_m, y_m), m = 1, 2, ..., M$ are fulcrums, $x$ is the state, $y$ is the likelihood, and $f(\cdot)$ is the required Li-PDF.

The dependence of the likelihood function on the parameters can be either linear or nonlinear. For the nonlinear likelihood function, solutions include approximate linearization with tolerable errors (given in appendix) and conversion methods of the nonlinearity (see subsection 3.5). Otherwise, some nonlinear regression method is required, such as the Gauss-Newton method. The Gauss-Newton method is one algorithm for minimizing the sum of the squares of the residuals between data and nonlinear equations, in which the least-squares theory may be used.

In what follows, we first consider the basic univariate variable fitting, while the intractable multivariate fitting will be described in subsection 3.4 where a local smoothing strategy is proposed for convenient implementation.

Normally, one will try to select a function $L(x)$ that depends linearly on the parameters, in the form of

$$L(x) = c_1\phi_1(x) + c_2\phi_2(x) + \ldots + c_k\phi_k(x) \qquad (20)$$

where $\{\Phi_i(x)\}$ are a priori selected sets of functions, for example, the set of monomials $\{x^{i-1}\}$ or the set of trigonometric functions $\{\sin\pi i x\}$, and $\{c_i\}$ are parameters which must be determined. In this paper, we call $k$ the order of the fitting function. In over-determined systems, as in our case, $k$ is much smaller than the number $M$ of fulcrums

$$M \gg k \qquad (21)$$

To specify the form of the functions in (20), the best case is when the function is known in advance. Otherwise, reasonable assumptions and offline searching for the optimal fitting model is necessary. To find the optimal fitting model, offline study might be helpful. Once the approximating function form and fulcrums have been defined, as explained in sections 4.2 and 4.3 respectively, the next step is to determine the population parameters $c_1, c_2, \ldots, c_k$ to get a "good" approximation. As a general idea, the residuals

$$d_m = f_m - L(x_m; c_1, c_2 \ldots, c_k), \quad m = 1, 2, \ldots M \qquad (22)$$

are simultaneously made as small as possible. One tries to make some norm of the $M$-vector d =

$[d_1, d_2, ..., d_M]^T$ as small as possible - typically such as the 2-norm

$$\|d\|_2 = \left(\sum_{m=1}^{M} |d_m|^2\right)^{1/2} \qquad (23)$$

This leads to a linear system of equations to determine the minimum $\hat{c}_k$'s. The resulting approximation $L(x; \hat{c}_1, \hat{c}_2, ..., \hat{c}_k)$ is known as the least squares approximation to the given data and $\hat{c}_k$'s are called least squares estimates of the population parameters.

An appropriate fitting model and proper distributed fulcrums are two critical factors needed to achieve good fitting results. The Goodness-of-fit could be tested to decide whether it is possible to proceed or search for a more suitable Li-PDF model, one that will better represent the true observation. Available Goodness-of-fit tests include the Kolmogorov-Smirnov test, Anderson-Darling test, Chi-Square test, etc. [41].

*3.4 Piecewise fitting function and Kernel density estimation*

For many practical systems, however, it is difficult or even impossible to find a single function to represent the likelihood function in the entire state space, especially for the intractable multivariate fitting (*Hyper-surface problem*). As such, a flexible piecewise constant form could be chosen where the fitting function is of lower order. Accompanied with the piecewise fitting/local regression strategy, the linearization of the nonlinear dependence on parameters will be more theoretically tenable and easier to implement. This can also reduce the required fitting function order that promises a smaller linearization error (see also the appendix).

To perform the *Piecewise/Segmented fitting*, the independent variable is partitioned into intervals, and then a separate segment is fitted to each interval and the boundaries between the segments. The fitting function is thus a sequence of grafted sub functions

$$L(x;C) = F_1(x;C_1) \quad x_1 \leq x \leq x_2$$
$$= F_2(x;C_2) \quad x_2 \leq x \leq x_3 \qquad (24)$$
$$\vdots$$
$$= F_r(x;C_r) \quad x_r \leq x \leq x_{r+1}$$

where $x_1, x_2, \ldots, x_r$ are called join points which are boundaries between intervals. In our current approach, sub functions $F_i(x; C_i)$ are of the same order $k$, the number of fulcrums $M_i$ in each interval (including two join points) satisfies

$$\forall i \in [1, r]: M_i > k+1 \qquad (25)$$

It is shown in [40] that the piecewise constant approximations are the best when the densities are reasonably smooth in the scale of the grid. This indicates that the piecewise intervals should be partitioned such that the likelihood PDF in each interval is reasonably lower-order smooth.

The employment of numerical fitting will potentially slow down the weight concentration of particles as it reduces the high likelihood but increases the low likelihood. This will be helpful to alleviate the sample degeneracy/impoverishment, which is a particular challenge for the particle filter [32].

We should note that our goal is to calculate the likelihood of particles but not to explicitly obtain the Li-PDF, which is only an intermediate process. Thus, in the piecewise fitting, we can use nonparametric local smoothing techniques, e.g. Kernel Density Estimator (KDE), to derive the likelihood of particles by using the fulcrums without explicitly obtaining the Li-PDF. This method, termed the implicit Li-PDF approach, will greatly simplify the multivariate fitting for convenient implementation. Next, we will illustrate how it works. For a particle with state $x_t^{(j)}$, denoting its nearest $M_j$ fulcrums in a limited scale and their likelihoods as $\{x_i, p_i\}_{i=1,2,\ldots,M_j}$, the required likelihood $p(y_t^{(j)}|x_t^{(j)})$ KDE can be defined as the Nadaraya-Watson kernel-weighted average of these fulcrum likelihoods

$$p\left(y_t^{(j)} \middle| x_t^{(j)}\right) = \frac{\sum_{i=1}^{M_j} K_{h_\lambda}\left(x_t^{(j)}, x_i\right) p_i}{\sum_{i=1}^{M_j} K_{h_\lambda}\left(x_t^{(j)}, x_i\right)} \qquad (26)$$

with the kernel given as

$$K_{h_\lambda}\left(x_t^{(j)}, x_i\right) = D\left(\frac{\left\|x_t^{(j)} - x_i\right\|}{h_\lambda\left(x_t^{(j)}\right)}\right) \qquad (27)$$

where $h_\lambda$ is a specified bandwidth termed the interval width, and $D(t)$ is a positive real valued function whose value does not increase with an increasing distance between $x_t^{(j)}$ and $x_i$.

Two quite convenient kernel smoothers are available: the nearest neighbor smoother and the uniform kernel average smoother. The idea of the nearest neighbor (NN) smoother is as follows. For each point $x_t^{(j)}$, take $M_j$ nearest neighbor fulcrums $x_i$ and estimate the likelihood of the particle $p(y_t^{(j)}|x_t^{(j)})$ by averaging the values of these neighbors likelihood. Formally, for (27)

$$K_{NN}\left(x_t^{(j)}, x_i\right) = h \qquad (28)$$

In contrast to this, the uniform kernel function can be defined as

$$K_{uniform}\left(x_t^{(j)}, x_i\right) = \frac{h}{\left\|x_t^{(j)} - x_i\right\|} \qquad (29)$$

As the estimate of this uniform kernel smoother, every fulcrum $x_i$ in the bounded interval $h$ contributes to the likelihood of the particle $x_t^{(j)}$ in a manner inversely proportional to their distance from the particle. The NN smoother and uniform kernel smoother will be applied subsequently in our simulations in Section 4.

It is necessary to note that the observation functions are actually limited to a few simple types in real life filtering problems. For example, in tracking and localization problems, the observation functions that correspond to different types of sensors, whether detecting the position or bearing of the target (s), are simply of a lower order than 3. Furthermore, the observation function is often assumed to be with additive Gaussian noises. That is to say, the observation function and the likelihood function are not arbitrary in math; instead, they are a smooth distribution with low order, which are much easier for numerical fitting than they seem.

*3.5 A polynomial fitting example*

While boosting the processing speed of PF, it is important for our approach to guarantee the approximation accuracy. In order to have an intuitive understanding of the numerical fitting process and its results, the following popular univariate SSM is considered. The system dynamic and observation equations are, respectively,

$$x_t = \frac{x_{t-1}}{2} + \frac{25x_{t-1}}{1+x_{t-1}^2} + 8\cos(1.2(t-1)) + u_t \qquad (30)$$

$$y_t = 0.05x_t^2 + v_t \qquad (31)$$

where $u_t$ and $v_t$ are zero mean Gaussian random variables with variance 10 and 1 respectively.

Assuming the unknown observation equation is $y_t^{(i)} = g(x_t^{(i)})$, the likelihood function can be obtained through one more step, i.e. the following Gaussian model

$$L_t(y_t^{(i)}) = \frac{1}{\sqrt{2\pi}} \exp\left(-\frac{1}{2}(y_t - y_t^{(i)})^2\right) \qquad (32)$$

where $y_t$ is the real observations and $y_t^{(i)}$ is the observation of particle/fulcrum $x_t^{(i)}$. As shown, the likelihood function is in fact nonlinear (most commonly it is an exponential function which corresponds to Gaussian observation noise). Instead of using nonlinear fitting methods that are sometimes quite complex to implement, one may choose to linearize the nonlinearity. Equation (32) can be linearized by taking its natural logarithm to yield

$$\ln L_t(y_t^{(i)}) = \ln\frac{1}{\sqrt{2\pi}} \times \left(-\frac{1}{2}(y_t - y_t^{(i)})^2\right) \qquad (33)$$

Thus, the function $\ln L_t(x)$ with independent variable $x$ has a linear dependence on the parameters. However, for this model we only fit the observation function $y_t^{(i)} = g(x_t^{(i)})$. Fulcrums can be uniformly distributed with parameter $r = 1$ in (11), and the 2-order polynomial in the following trinomial form is assumed as the observation equation

$$y = c_3 x^2 + c_2 x + c_1 \qquad (34)$$

Then we get the Li-PDF$(L_t °g)(x_t^{(i)})$, which is

$$L_t(x_t^{(i)}) = \frac{1}{\sqrt{2\pi}} \exp\left(-\frac{1}{2}(y_t - c_3 x_t^{(i)2} - c_2 x_t^{(i)} - c_1)^2\right) \qquad (35)$$

This is known as a nonlinear conversion that changes the nonlinear fitting function to a linear one, which has high potential to apply to SSM with Gaussian observation noise. In Fig. 2, the 'ideal' true observations ($y = 0.05x^2$) without noise are shown in 'black' curve and its direct noisy observations in (31) of 100 random samples are shown in red circles. The least squares fitting results of the noisy observations in (34) of $M$ fulcrums ($M = 5, 10, 30, 50$ separately; for 2-order function, $M \gg 2$ according to the condition (21)) are shown with respective colored lines. The fitted functions (in one trial) are as follows:

$y = 0.0478x^2 + 0.0140x + 0.7750$ (5 fulcrums)
$y = 0.0398x^2 + 0.0122x + 0.8667$ (10 fulcrums)
$y = 0.0486x^2 - 0.0121x + 0.2818$ (30 fulcrums)
$y = 0.0504x^2 + 0.0245x - 0.1050$ (50 fulcrums)

The results show that our numerical fitting approach gets more accurate observations than direct observation. These good results benefit from our pre-knowledge that the observation equation is a 2-order polynomial, although this is a fairly weak assumption. Obviously, the fitting results can grow stronger as the knowledge of the observation model improves. This will be further shown in our simulation section 4.1. For example, if we know the fitting function is in the monomial form

$$y = c_3 x^2 \qquad (36)$$

then more precise fitting results can be obtained (as shown in Fig. 3) by using the same fulcrums. For example, in one trial, we get: $c_3$=0.0570 (5 fulcrums), $c_3$=0.0518 (10 fulcrums), $c_3$=0.0530 (30 fulcrums), $c_3$=0.0487 (50 fulcrums).

The above example has just exhibited a potential application of the numerical fitting tool to estimate the observation function (if it is unknown and needs to be estimated). Furthermore, if the observation equation is known (as the case of most common SSMs), the numerical fitting method can be applied in a batch manner in which it does not need to online fit the equation of (34) or (36) for each step but directly use the known one. The fitted observation function or even the exact function $y = 0.05x^2$ can be taken as granted in the subsequent filtering steps to infer the likelihood of particles for saving computation. We refer to this form of fitting method as the batch fitting in which the function of interest is constant. This however is inapplicable to the likelihood PDF which is generally not a constant function. We will demonstrate this in our simulations.

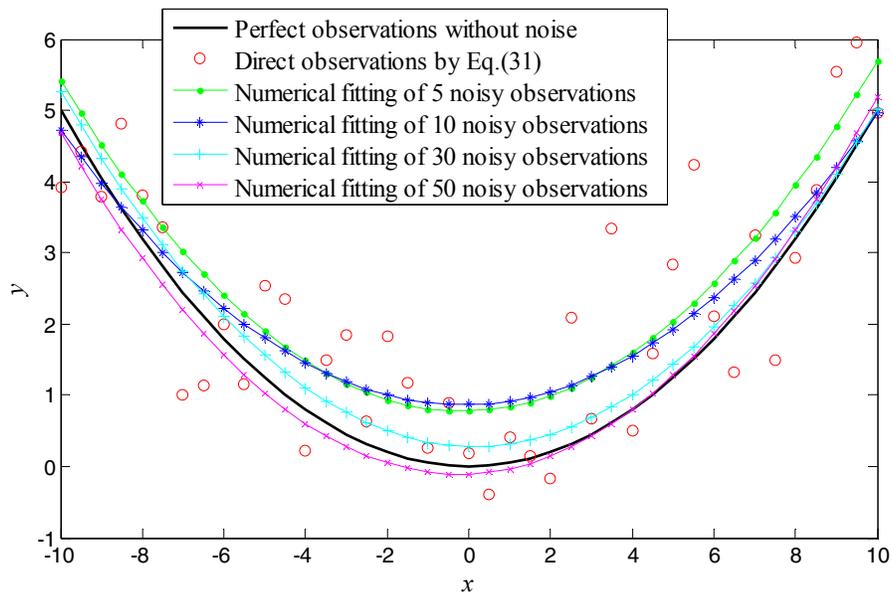

Fig. 2 Observations without noise, observations with noise and Eq. (34)-based fitting function using different number of fulcrums

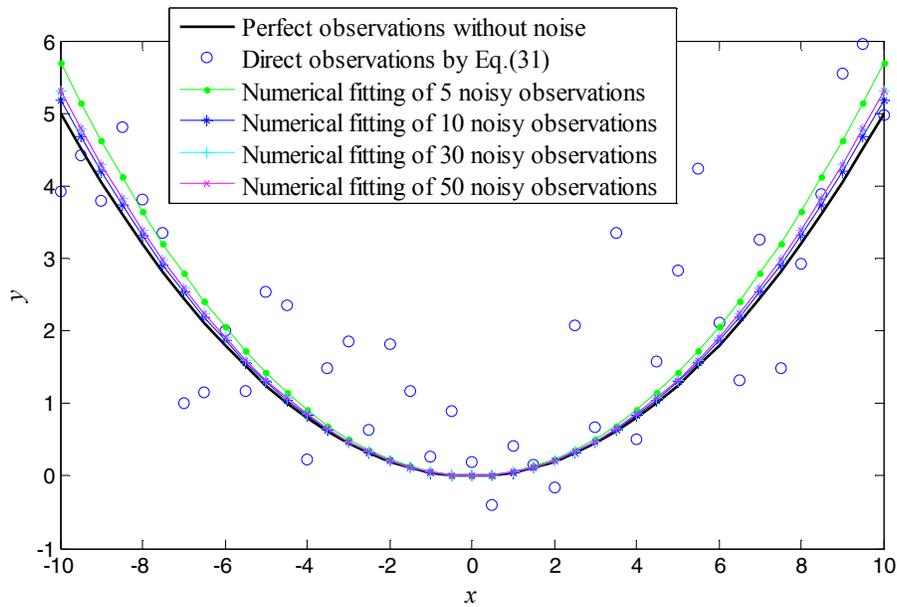

Fig. 3 Observations without noise, observations with noise and Eq. (36)-based fitting function using different numbers of fulcrums

## 4 SIMULATIONS

In order to verify the validity of our approach, typical filtering problems including the aforementioned 1-dimensional model and robot localization are considered in this section.

### 4.1 One-dimensional model: Fitting observation function

For the nonlinear system described in equations (30), (31), the root mean square error (RMSE) in the time series is used to evaluate the estimation accuracy, which is calculated by

$$\text{RMSE} = \left(\frac{1}{T}\sum_{t=1}^{T}(x_t - \hat{x}_t)^2\right)^{1/2} \qquad (37)$$

where $\hat{x}$ is the estimate of the state, $T$ is the sum of iterations. A big $T = 10{,}000$ is chosen and a sequence number of particles from 10 to 500 with interval of 10 is separately used.

First, different orders of polynomial with 10 fulcrums are used in the regression model (34) to fit the observation function. The RMSE results of the Li-PDF based PFs are given in Fig. 4, from which it can be seen that the 1st-order polynomial fitting result is really poor whereas 2nd-order and higher form polynomials get much better estimation accuracy. This indicates that a proper (no smaller than the true order) fitting function is critically important for our approach.

Secondly, the RMSE of the basic particle filter and the Li-PDF based particle filters using different numbers of fulcrums (for 2nd-order fitting polynomial) is given in Fig. 5. The results indicate that the Li-PDF based PF can obtain a comparable estimation accuracy with the SIR filter, as long as an enough number of fulcrums are used (satisfying Eq. (13)).

Thirdly, the comparison of the Li-PDF PF (including the batch form with given observation function) and several other known nonlinear filters including the SIR PF, auxiliary PF (APF) [34], Gaussian PF (GPF) [26], Kernel PF (KPF) [36] and Unscented Kalman filter (UKF, with the unscented transform parameter set as $\alpha = 1, \beta = 0, \kappa = 2$) [44] are given in Fig. 6 and Fig. 7 for the RMSE and processing time respectively. Here, the Li-PDF PF uses 10 fulcrums and fitting function (36) at all steps, while the batch Li-PDF PF uses 100 fulcrums and fitting function (36) at the first step. Their average performance over different numbers of particles is given in Table I where the weight updating is not vectorized in Matlab as explained below.

The results show that the UKF does not work for this highly nonlinear model as its RMSE is high (see [26] as well) while all PFs perform very similarly. In detail, the GPF and the APF is somewhat inferior to the (batch) Li-PDF PFs, the SIR and the KPF on average. When the number of particles is small, the (batch) Li-PDF PFs performs better than others.

The Matlab programming itself can highly determine the computing speed. Especially, the vectorization i.e. the data are processed in the unit of matrix can highly speed up the computation. The processing time of all filters are given in Fig.7 for the case without vectorization of the weight updating step and are given in Fig.8 for the case of using vectorization for updating. As a fact, in the multi-dimensional state space, vectorization might be infeasible (e.g. when dimensions are correlated), i.e. Fig.7. We must therefore deal with each particle separately, the computational demand of the PFs (including GPF, KPF, APF and SIR) will unsurprisingly increase in proportion with the growth of the number of particles used as shown in Fig.7; see our next simulation. For this case, the UKF is obviously the fastest. The batch Li-PDF based PF is the second (it does not need to online fit the equation of (36) for each step but directly use the known one), GPF is the third (no resampling is needed) and the Li-PDF PF is the fourth (the computation does not linearly depend on the number of particles). All of them are highly faster than the KPF, APF and SIR (especially when the number of particles is large). KPF and APF are slower than the SIR. In contrast, in the Li-PDF PFs the number of fulcrums is constant and their likelihood calculation does not necessarily increase with the number of particles. This is just the superiority of our approach that can remove the high dependence of the computational time required on the number of particles used. What has been increased is only the inference calculation of the likelihood of particles via the likelihood of fulcrums which is computationally fast.

To note, the processing speed of the Li-PDF PF can only be improved when the time cost for the fitting is less than the likelihood computation it has saved. Since the updating step (31) when the vectorization is employed is nothing more than the job of solving (34), it is not surprising that the computing speed of the Li-PDF PF has not been improved but instead reduced in such a simple simulation when vectorization is utilized or when the number of particles is very small. As noted, this model is significantly different to the applications of multiple dimensionalities, in which the weight updating is much more computationally intensive than the state prediction and resampling and in which the multi-dimensional state is often unable for the vectorization of the weight updating.

In particular, in such a 1D model, the resampling will not take a large part of the computation in the PF and therefore the GPF is very fast which does not need to perform resampling. However, in high-dimension models where the weight updating is much more computation-consuming than the resampling and the state prediction, the computation advantage of the GPF will not be so obvious but the Li-PDF will further speed up the PF. This will be demonstrated in our next simulation.

Table I Average Performance of filters (without vectorization)

|              | N=50  |        | N=200 |        |
|--------------|-------|--------|-------|--------|
|              | RMSE  | Time   | RMSE  | Time   |
| UKF          | 7.641 | 3.491  | 7.641 | 3.427  |
| SIR          | 5.623 | 10.840 | 4.895 | 38.196 |
| APF          | 5.805 | 18.044 | 5.060 | 66.885 |
| KPF          | 5.706 | 12.514 | 4.896 | 42.910 |
| GPF          | 5.821 | 2.412  | 5.038 | 5.955  |
| Li-PDF       | 5.546 | 12.210 | 5.004 | 13.906 |
| Batch Li-PDF | 5.548 | 1.704  | 4.883 | 3.170  |

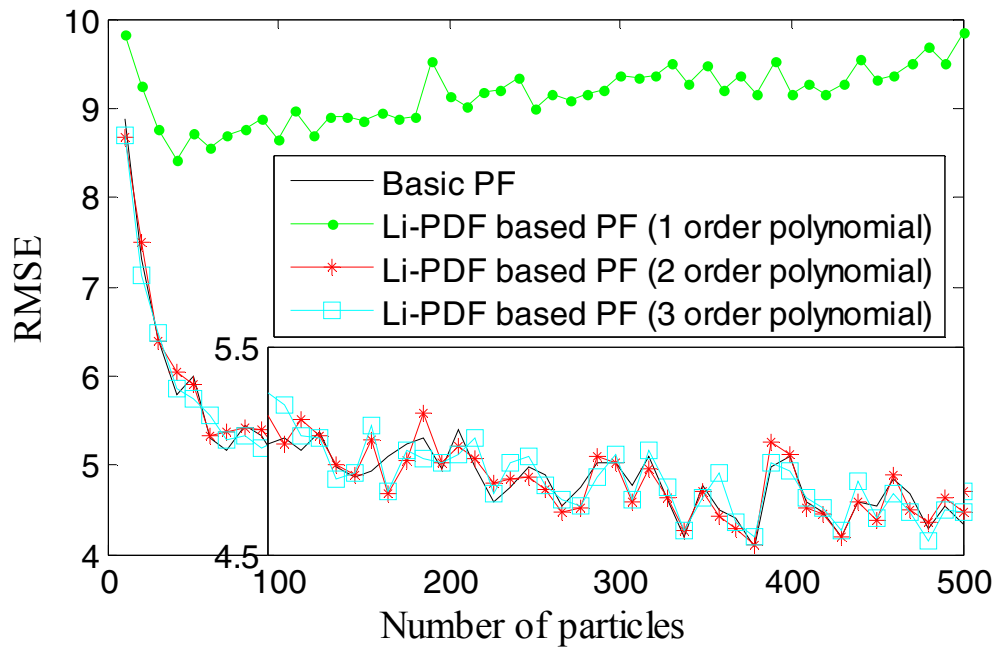

Fig.4  RMSE of the basic SIR PF and the Li-PDF based PFs that use different orders of polynomials

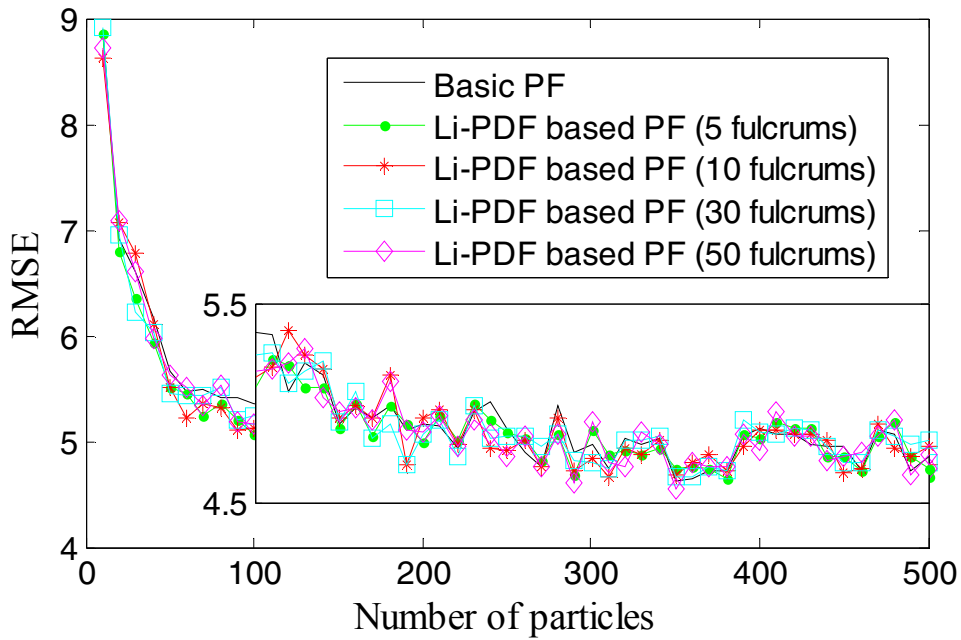

Fig.5  RMSE of the basic PF and the Li-PDF based PFs that use different numbers of fulcrums

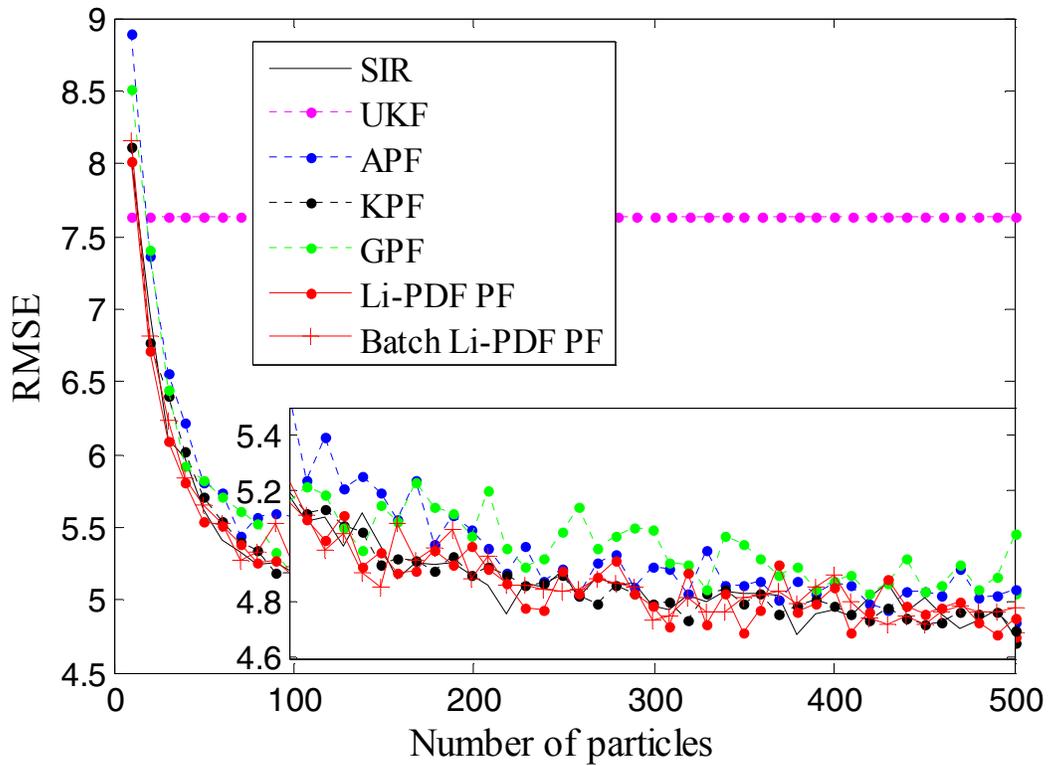

Fig.6 RMSE of different filters against the number of particles used

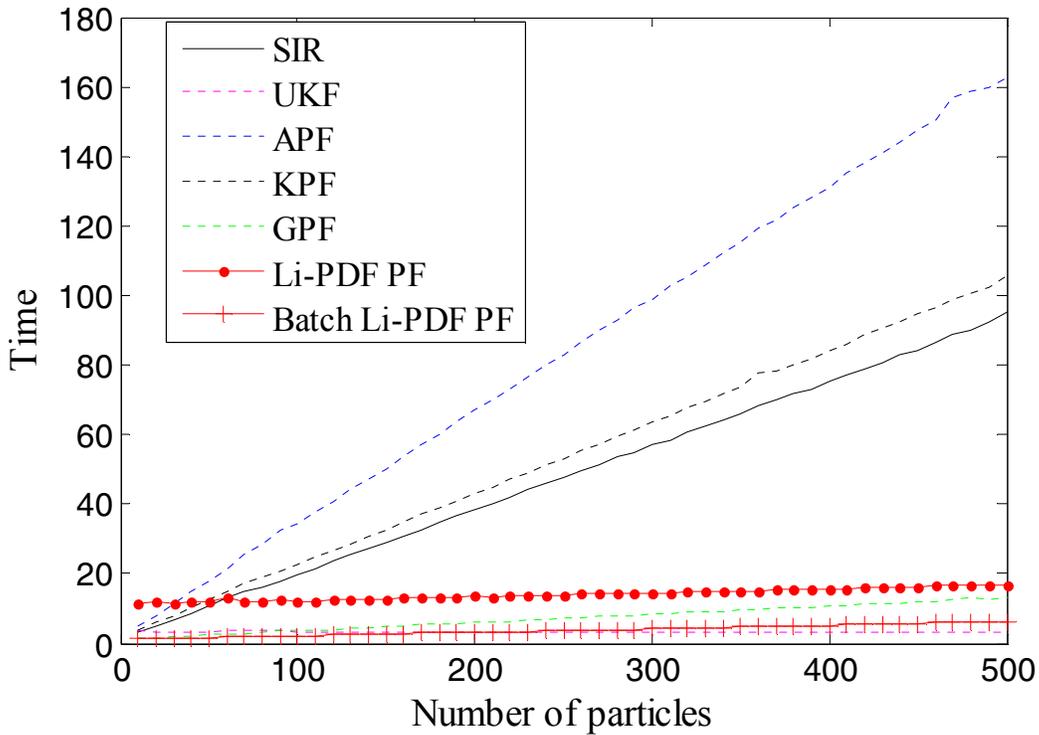

Fig.7 Computing time of different filters for 10,000 steps (the weight updating of particles is not vectorized)

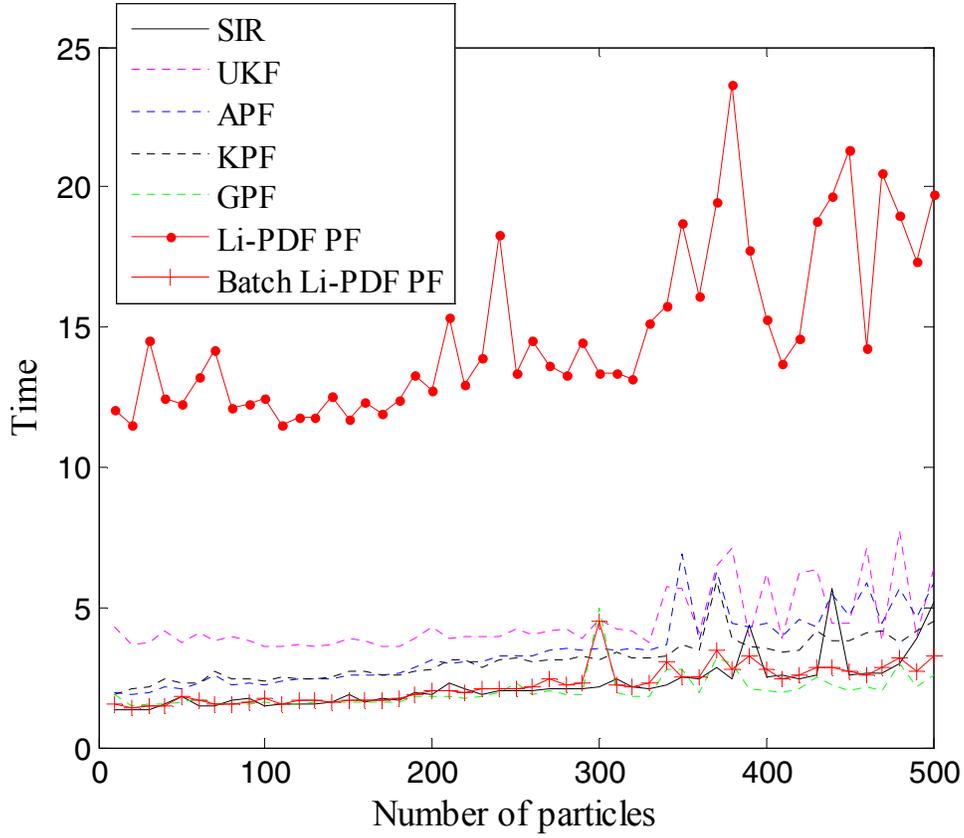

Fig.8 Computing time of different filters for 10,000 steps (the weight updating of particles is vectorized)

*4.2 Multi-dimensional model: Fitting likelihood PDF*

In general, the SIR PF seems to be the fastest computationally while existing variant PFs are often more computationally intensive except for the GPF which does not need to perform resampling. Therefore, in our second simulation, our comparison has not included all of these advanced PFs except SIR and GPF.

The application of PF to mobile robot localization is usually referred to as the Monte Carlo localization (MCL) [14]. In this section, we present a typical MCL application via simulation for accurate comparison and analysis. In order to develop the details of MCL, let $x_t = (x, y, \theta)^T$ be the robot's position in Cartesian space $(x, y)$ along with its heading direction $\theta$ denotes the robot's state at time instant $t$, $y_t$ is the observation at time $t$, and $u_t$ is the odometer data (control observation) between time $t-1$ and $t$. The prediction and observation updating model required by MCL is set as follows

$$p(x_t | y_{t-1}, u_{t-1}) = p(x_t | x_{t-1}, u_{t-1}) \times p(x_{t-1} | y_{t-1}, u_{t-2}) \quad (38)$$

$$p(x_t|y_t, u_{t-1}) = \frac{p(y_t|x_t) p(x_t|y_{t-1}, u_{t-1})}{p(y_t|y_{t-1}, u_{t-1})} \quad (39)$$

Supposing $u_{t-1}$ has a movement effect $(\Delta s, \Delta \theta)^T$ on the robot, $\Delta s$ is the translational distance, and $\Delta \theta$ the change of robot's heading direction from time $t-1$ to $t$. Then, the motion model $p(x_t|x_{t-1}, u_{t-1})$ can be easily obtained as

$$x_t = x_{t-1} + \begin{bmatrix} \cos\theta_{t-1} & 0 \\ \sin\theta_{t-1} & 0 \\ 0 & 1 \end{bmatrix} \begin{bmatrix} \Delta s \\ \Delta \theta \end{bmatrix} + v_{t-1} \quad (40)$$

where $v_{t-1}$ is the system uniform noise with zero mean and $\mathrm{diag}[\Delta s \times 20\%, \Delta \theta \times 5\%]^T$ variance in our case. In particular, the particle which falls into obstructs will be discarded (by setting its weight to zero) in resampling.

The likelihood-based weight updating $p(y_t|x_t)$ depends on the perceptual data, which can be proximity data (scanning radar or sonar), or visual data from image/video, which can be much more complex and computationally intensive. The nearest-neighbor data association used for scan matching in our simulation can be described as

$$p(y_t|x_t) \propto \frac{1}{\sqrt{(2\pi)^n |S_{i,j}|}} \exp\left(-\frac{1}{2}(\hat{y}_i - y_j)^T S_{i,j}^{-1}(\hat{y}_i - y_j)\right) \quad (41)$$

where $S_{i,j}$ is the covariance matrix of the difference $\hat{y}_i - y_j$, the Gaussian observation noise is $s \sim \mathcal{N}(0,5)$ for each scanning distance, $n$ is the number of scanning lines and we choose $n = 36$ and $180$ respectively in our case. A bigger $n$ indicates a higher resolution and is more time-consuming. Equation (41) also indicates it is impossible to apply vectorization for weight updating as is done in the one-dimensional SSM like (30~31).

The Li-PDFs could facilitate the entire state space $(x, y, \theta)^T$ or simplify the Cartesian space $(x, y)^T$ only. This simplification is possible because the direction $\theta$ relies strongly on its position $(x, y)^T$ if the covariance matrix $S_{i,j}$ can be known based on the observations. In this case, it is set $r^{[l]} = 1$ in (11), $p = 10$ in (17) so that 100 fulcrums are used in the position space $(x, y)^T$ and the linear uniform fitting method is adopted. In addition, the Li-PDF approach is applied after $t = 3$ since at the starting stage ($t \leq 3$) particles are very widely distributed

(e.g. the case is plotted in Fig. 9 for $t = 2$) which is unsuitable for constructing the Li-PDF. To determine whether it becomes suitable for fitting, one suggestion here is to measure the variance of the particle distribution. Here, we use the threshold method.

To evaluate the filtering performance, the Euclidean Distance (ED) is defined between the position estimate $(\hat{x}, \hat{y})^T$ and the ground truth $(x, y)^T$ that is calculated by

$$ED = \sqrt{(x - \hat{x})^2 + (y - \hat{y})^2} \qquad (42)$$

The path of the robot is from 'S' to 'T' in Fig. 9. The points represent particles and the rectangle boxes represent the stops of robot. In one trial when 500 particles and 100 fulcrums are used and it is set $n=36$, the distribution of particles, surfaces of Li-PDF, and the discrepancy between likelihood of particles that are obtained by the Li-PDF and by direct observation-based calculation at two stops are given in Fig. 10.

Both the number $N$ of particles and the number $M$ of fulcrums are critical to the performance of PFs. To capture the average performance, 100 MC trials were run. The EDs when different numbers $N$ of particles are used are plotted by time steps in Fig. 11, which indicates that the Li-PDF approach has indeed reduced the estimation accuracy somewhat as compared with the SIR PF and the GPF (the latter two perform similarly). In view of the high nonlinearity of the simulation model, the estimation accuracy is acceptable. For a huge number of particles e.g. 500, a small number of fulcrums (100) can fit the likelihood efficiently. Furthermore, it can be seen that for the same number of fulcrums, applying more particles does not generate better results. This also exposes the fact that it is not suitable to use a relatively small number of fulcrums to fit the likelihood of too many particles; instead, the ratio of the number of fulcrums to the number of particles should be set in a reasonable scope. Too small a ratio (too few fulcrums) leads to worse results, while too high a ratio does not benefit the processing speed.

The mean ED from stops 3 to 24 against the number $M$ of fulcrums is plotted in Fig. 12, which shows that the larger the number of fulcrums, the more accurate is the approximation. The reason that the result is better for a larger number of particles is because of the very first steps as shown in Fig. 11 (while in the latter steps, a greater number of particles leads to worse estimation accuracy). Note that the Li-PDF approach does not fit during the initial stage where particles are distributed broadly in the state space. The computing time of the Li-PDF based PF, SIR and GPF against the number of particles are given in Table II and Table III respectively for the scanning data size $n = 36$ and 180. The results demonstrate again the fast processing advantage of our Li-PDF approaches (that the computing speed of PF is no more limited so heavily by the number of particles), especially when informational-rich observations ($n = 180$) are applied.

The resampling that is dimension-free takes a large part of computation in the simple 1D model but not in this complicated multi-dimensional models. Here, the weight updating of particles is the primary computation of the PF. Therefore, the GPF is faster than the SIR (not so obviously as in the first 1D simulation) but is much slower than the Li-PDF MCL. Comparably, the Li-PDF approach is qualified to significantly reduce the computation requirement while maintaining approximation quality.

There is a trade-off between increasing the processing speed by reducing likelihood calculation and improving the estimation accuracy by maintaining accurate likelihood calculation. As such, it is highly recommended to use an off-line search for the optimal number of fulcrums, as well as the fitting function as aforementioned. The choice also depends on the practitioner's preference between the estimation accuracy and the processing speed. Our Li-PDF approach provides a choice for applications in which a fast processing speed is much preferred.

Table II Real-time Performance of PFs (*Second*) When the Scanning Data Size $n$=36

| Number of particles | 100 | 500 | 1000 |
|---|---|---|---|
| Basic SIR PF | 0.593 | 3.668 | 10.078 |
| Gaussian PF | 0.594 | 3.654 | 9.5165 |
| Li-PDF PF | 0.787 | 2.140 | 4.1419 |

Table III Real-time Performance of PFs (*Second*) When the Scanning Data Size $n$=180

| Number of particles | 100 | 500 | 1000 |
|---|---|---|---|
| Basic SIR PF | 3.417 | 20.937 | 50.768 |
| GPF | 3.566 | 15.707 | 47.321 |
| Li-PDF PF | 3.806 | 6.547 | 11.871 |

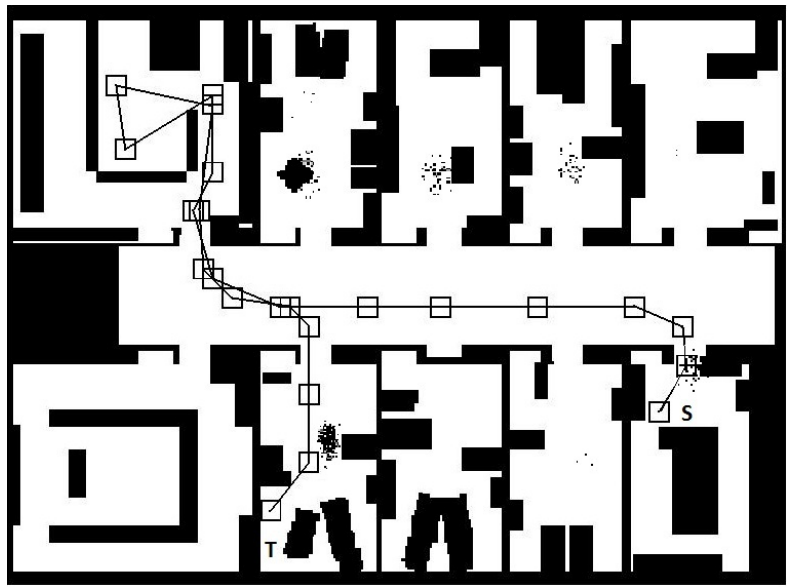

Fig.9 The robot trajectory and the distribution of particles (black point) when the robot is at the second stop

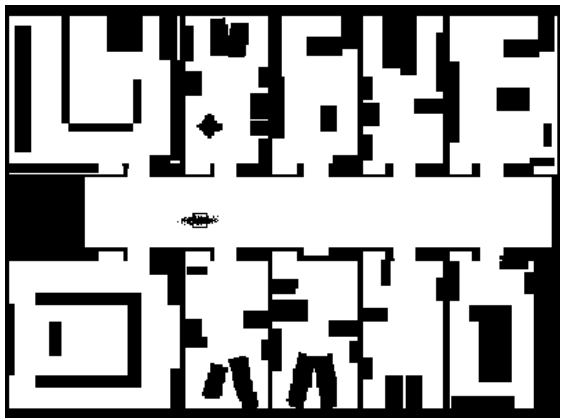
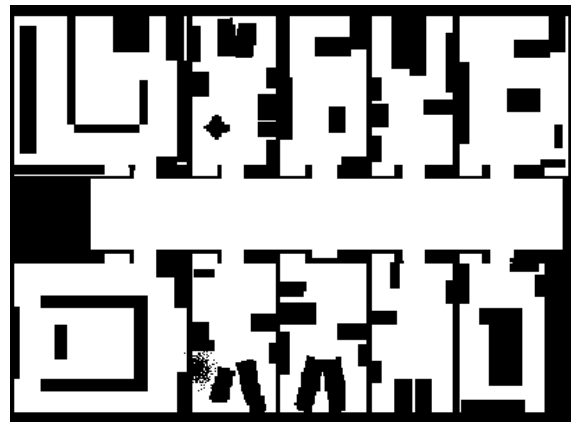

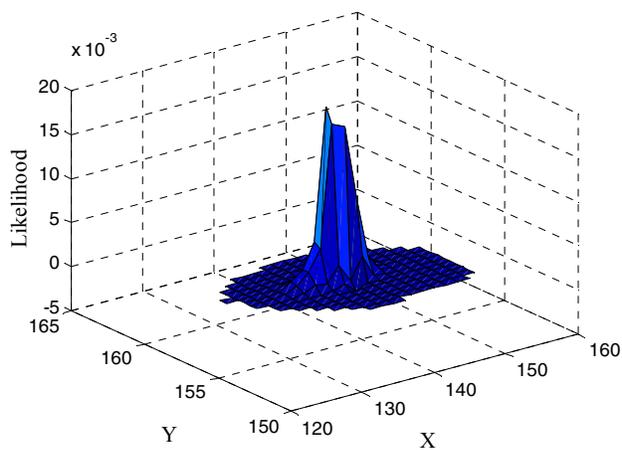
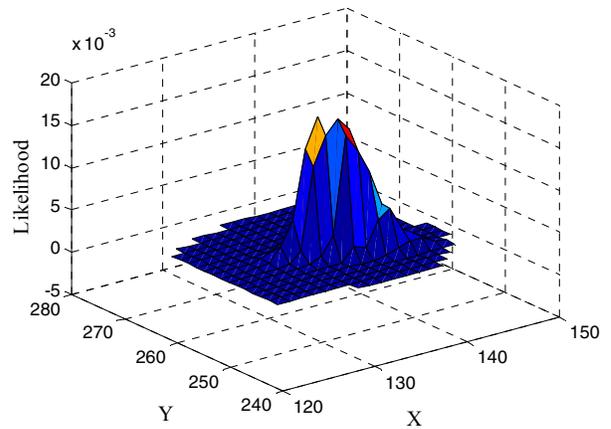

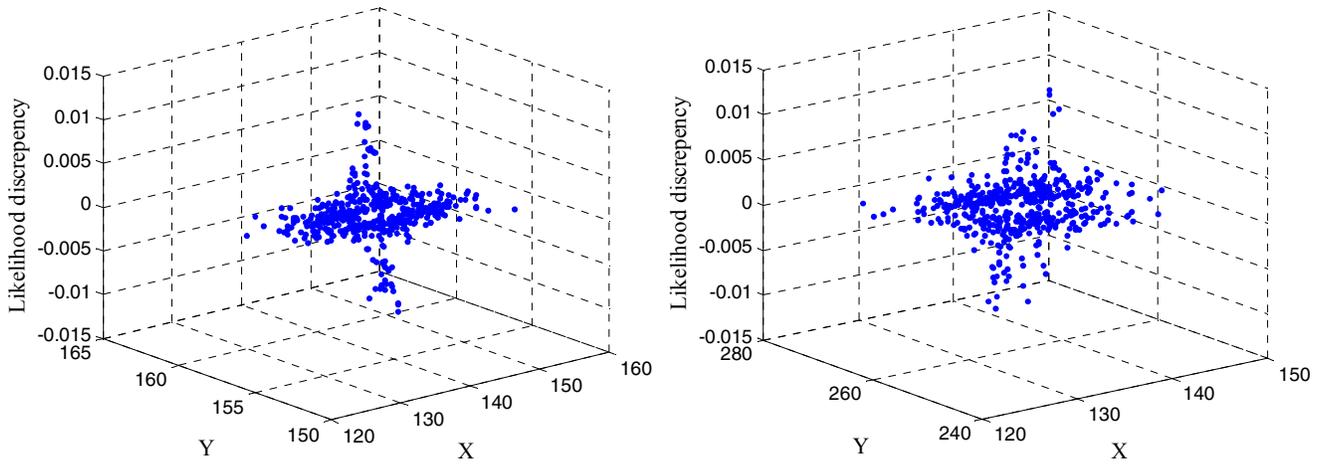

Fig.10 Distribution of particles (upper row), Li-PDF 3D surfaces (middle row) and the discrepancy between likelihood of particles that are obtained by the Li-PDF and by direct observation-based calculation (bottom row) in different stops ($N$=500, $M$=100, $n$=36).

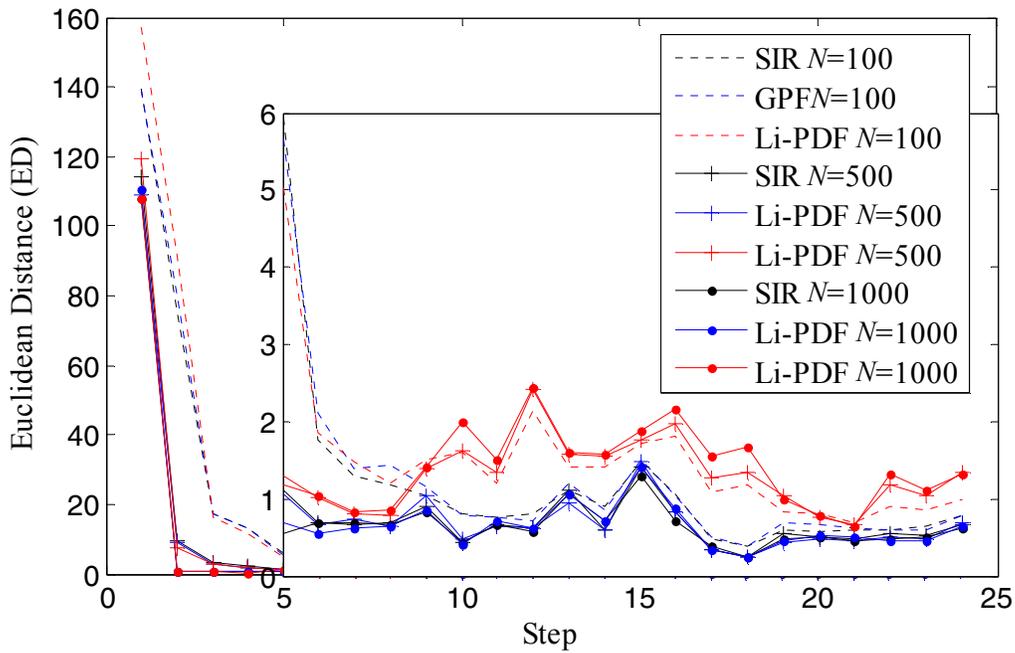

Fig.11 Estimation error by steps when different numbers of particles are used ($n$=36, $M$=100 in Li-PDF PFs)

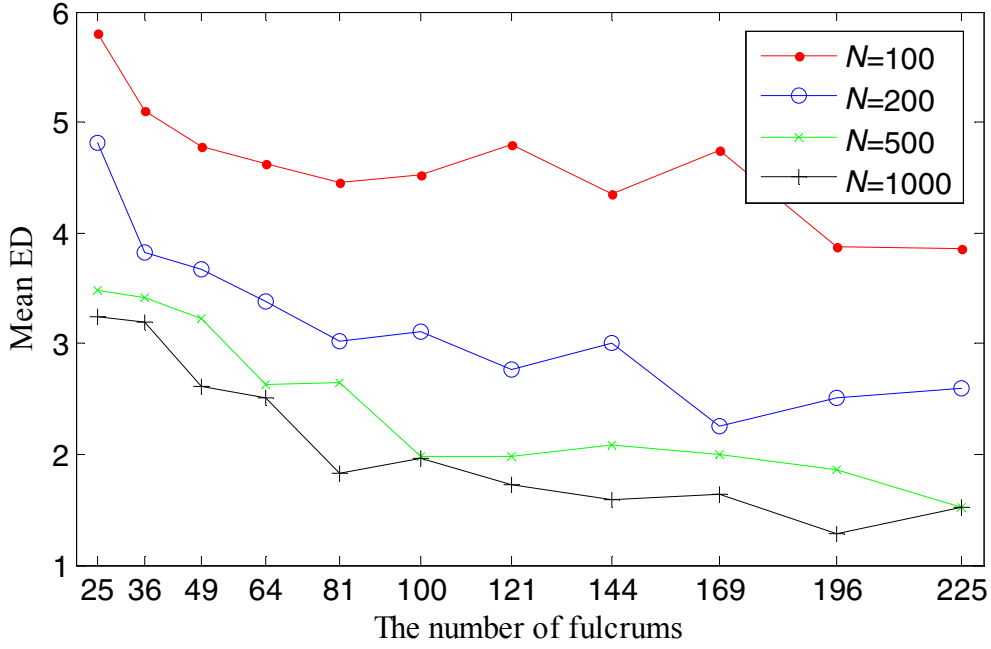

Fig.12 Mean ED against the number of fulcrums used ($n$=36)

*4.3 Color histograms based target tracking*

As a support document to our submission, the proposed approach in the submission is employed in particle filter to track a helicopter in a video as shown in Fig.13. To enhance the reproducibility of the experiment, we adopt a public instance shared by Sébastien Paris based on the very basic color histogram observation model proposed in [42] which is available on http://www.mathworks.com/matlabcentral/fileexchange/17960. The focus of our approach is not computer vision and for the detail of the observation model, see [1]. This experiment will not form part of the final manuscript.

We introduce the colour histograms based observation model given in [1] firstly. The colour distribution $p_y=\{p_y^{(u)}\}_{u=1,2,\ldots,m}$ at location $y$ is calculated as

$$p_y^{(u)} = f\sum_{i=1}^{I} k\left(\frac{\|y-x_i\|}{a}\right)\delta\left(h(x_i)-u\right) \qquad (43)$$

where $I$ is the number of pixels in the region ($I$=120 in our case), $\delta(\cdot)$ is the Kronecker delta function, $k(\cdot)$ is a weighting function of pixels, the parameter $a$ is used to adapt the size of the region, and $f$ is the normalization factor. It is satisfied that $\sum_{u=1}^{m} p_y^{(u)} = 1$.

$$k(r) = \begin{cases} 1-r^2 & r<1 \\ 0 & \text{otherwise} \end{cases} \quad (44)$$

$$a = \sqrt{H_x^2 + H_y^2} \quad (45)$$

$$f = \frac{1}{\sum_{i=1}^{I} k\left(\frac{\|y-x_i\|}{a}\right)} \quad (46)$$

where $H_x$, $H_y$ are the length of the half $x$, $y$ separate axes of the ellipse used to determine the color distribution.

We *initialize* the state space for the first frame manually. Each sample of the distribution represents an ellipse and is given as

$$s = \{x, \dot{x}, y, \dot{y}, H_x, H_y, \dot{a}\} \quad (47)$$

where $x$, $y$ specify the location of the ellipse, $\dot{x}$, $\dot{y}$ the motion, and $\dot{a}$ is the corresponding scale change. The fitting is currently implemented in the 2-dimesion position space

$$x_t = (x, y)_t \quad (48)$$

The *system dynamics* are described by a first-order auto-regressive model given as:

$$s_{t+1} = As_t + \mathcal{N}(0, R) \quad (49)$$

where $\mathcal{N}$ stands for Gaussian distribution, Matrix $A$ defines the deterministic component of the model and $R$ is the covariance as

$$A = \begin{pmatrix} 1 & \Delta t & 0 & 0 & 0 & 0 & 0 \\ 0 & 1 & 0 & 0 & 0 & 0 & 0 \\ 0 & 0 & 1 & \Delta t & 0 & 0 & 0 \\ 0 & 0 & 0 & 1 & 0 & 0 & 0 \\ 0 & 0 & 0 & 0 & 1 & 0 & 0 \\ 0 & 0 & 0 & 0 & 0 & 1 & 0 \\ 0 & 0 & 0 & 0 & 0 & 0 & 1 \end{pmatrix}, \quad R = \begin{pmatrix} R_{x_t} & 0 \\ 0 & R_e \end{pmatrix}$$

where $\Delta t = 0.7$ in our case as the tracking video is partitioned into 400 frames, $R_{xt}$ and $R_e$ are the position covariance and the ellipse covariance respectively.

$$R_{x_t} = \delta_{x_t} \begin{pmatrix} \Delta t^3/3 & \Delta t^2/2 & 0 & 0 \\ \Delta t^2/2 & \Delta t & 0 & 0 \\ 0 & 0 & \Delta t^3/3 & \Delta t^2/2 \\ 0 & 0 & \Delta t^2/2 & \Delta t \end{pmatrix}, \quad R_e = \begin{pmatrix} \delta_{H_x}^2 & 0 & 0 \\ 0 & \delta_{H_y}^2 & 0 \\ 0 & 0 & \delta_{H_\theta}^2 \end{pmatrix}$$

where, $\delta_{xt}^2 = 0.35$, $\delta_{Hx}^2 = 0.1$, $\delta_{Hy}^2 = 0.1$, $\delta_{H\theta}^2 = \pi/60$.

For the *observation model*, the likelihood of particles are proportional to the Bhattacharyya distance between the color window of the predicted location in the *n*th frame $p_n = \{p^{(u)}\}_{u=1,2,\ldots,m}$ and the reference $q = \{q^{(u)}\}_{u=1,2,\ldots,m}$

$$p(s_n) = \frac{1}{\sqrt{2\pi\sigma}} e^{-\frac{d^2}{2\sigma^2}} = \frac{1}{\sqrt{2\pi\sigma}} e^{-\frac{1-\rho[p,q]}{2\sigma^2}} \tag{50}$$

where $\sigma$ is the measurement noise $\sigma = 0.2$, $d$ is the Bhattacharyya distance defined as

$$d = \sqrt{1 - \rho[p,q]} \tag{51}$$

$\rho[p, q]$ is the Bhattacharyya coefficient. In our discrete case, the histograms are calculated in the HSV space using discrete 8×8×4 bins color window (*m*=256) as follows

$$\rho[p,q] = \sum_{u=1}^{m} \sqrt{p^{(u)} q^{(u)}} \tag{52}$$

It is obvious that the measurement updating function is much more computationally expensive than the state dynamic function.

In contrast experiments, the same number of particles is employed in the basic PF and the Li-PDF based PFs. For the Li-PDF approach, it is set $r^{[l]} = 5$ in (11), $p = 10$ in (17) of the submission so that 100 fulcrums are used. The Li-PDF is directly constructed by using the nearest neighbor based *griddata* fitting function in Matlab, which is applicable to the multiple dimension case. For most visual tracking video, we do not have the accurate data of the true trajectory but instead we can only have a coarse approximation of the ground truth by manually estimating frame by frame. The trajectories of our approach and basic PF when using 500 particles are given in Fig. 14 for one trial and the EDs of both filters against frames are given in Fig. 15. The mean ED at each step in the basic SIR PF is 3.8939 while in our approach it is 4.5209 based on the coarse ground truth. This indicates a reduction of the estimation accuracy in our Li-PDF approach. As the results therein revealed that the error of the Li-PDF approach is significant at the initial stage, the proposed Li-PDF approach is more suitable to be applied during the relative 'stable' tracking stage rather than at the beginning stage of the filter to achieve a better result.

Next, the amount of loss in tracking (the tracker completely drift), 5 out of 20 trials, is the same, indicating the comparable robustness in the proposed method comparing the basic PF. To display an insight of the Li-PDF, the Li-PDF surf is plotted in Fig. 16 and the discrepancy between the likelihood of particles that are obtained by the Li-PDF approach and by direct calculation is plotted in Fig. 17. Note that some regions/particles may be weighted negative in our approach as shown in Fig. 16. This makes sense for numerical fitting but not for the PF. To correct this, the negative weights can be set to zero.

The real-time performances of different filters are given in table CI. It can be seen that the computing time of the Li-PDF based PF does not linearly increase with the number of particles but the basic PF does. This demonstrates that the computational complexity of our approach is no longer heavily limited by the number of particles $N$ (but instead it depends more on the number $M$ of fulcrums used). This compares favourably with the computational cost of most current PFs if a small number of fulcrums are used. This will be appealing for the case that requires extremely massive number of particles. As stated, this is because the evolving system dynamics is computationally effortless as compared to the likelihood computation based on color histograms. This holds for most of visual tracking scene.

TABLE IV REAL-TIME PERFORMANCE OF PARTICLE FILTERS (*SECOND*)

| Number of particles | 100    | 250    | 500    | 1000   | 2000   |
|---------------------|--------|--------|--------|--------|--------|
| Basic PF            | 14.652 | 19.677 | 26.541 | 44.166 | 75.213 |
| Li-PDF PF           | 15.346 | 15.578 | 16.546 | 17.131 | 17.869 |

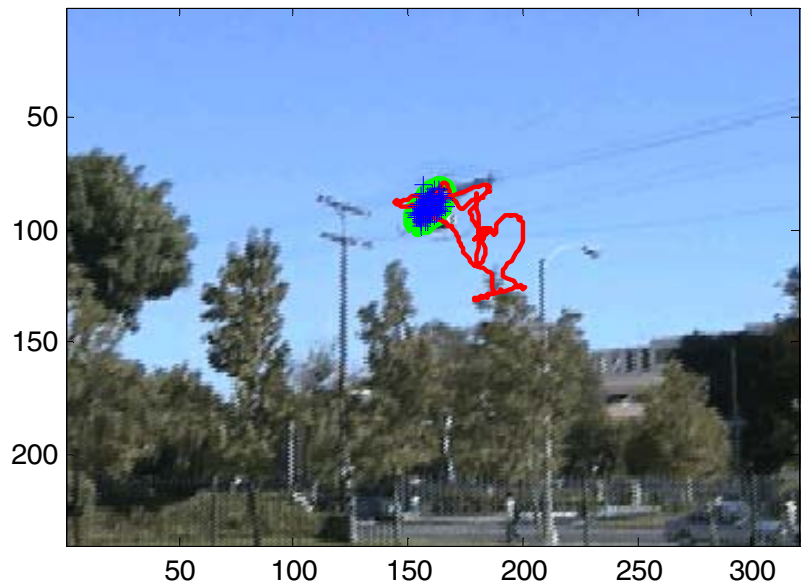

Fig.13 Snapshot of the last frame (red curve represents the estimated trajectory)

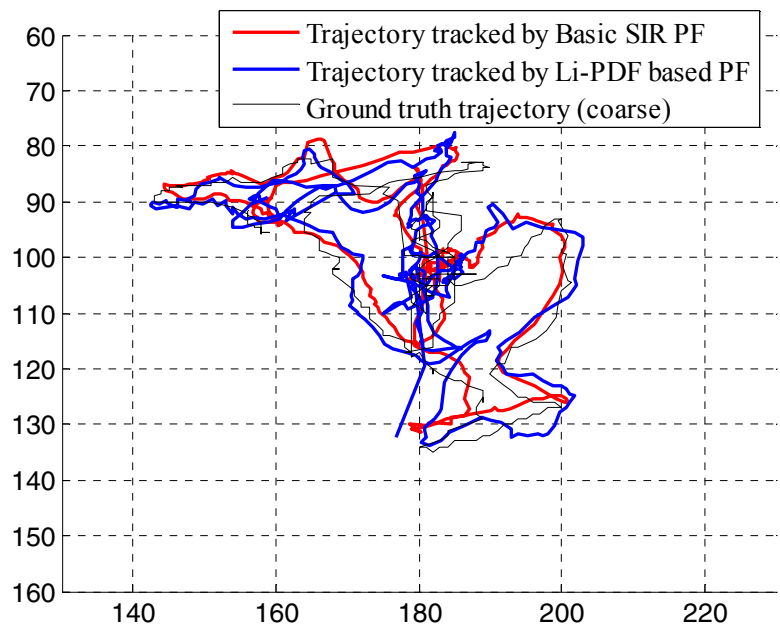

Fig.14 Helicopter trajectories comparison of the basic SIR PF and the Li-PDF based PF (the so-called ground truth trajectory is not accurate but is manually estimated only)

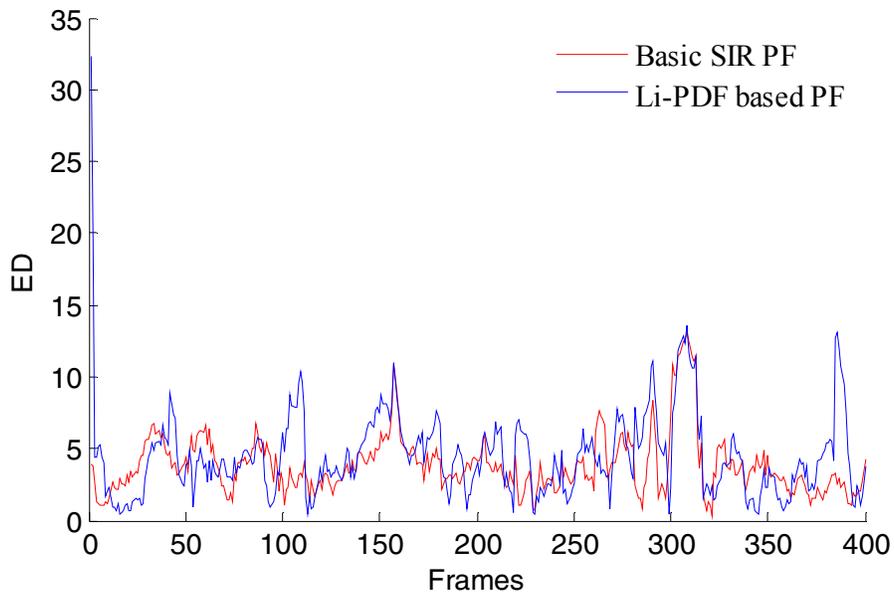

Fig.15 EDs of the basic SIR PF and the Li-PDF based PF based on a coarse ground truth

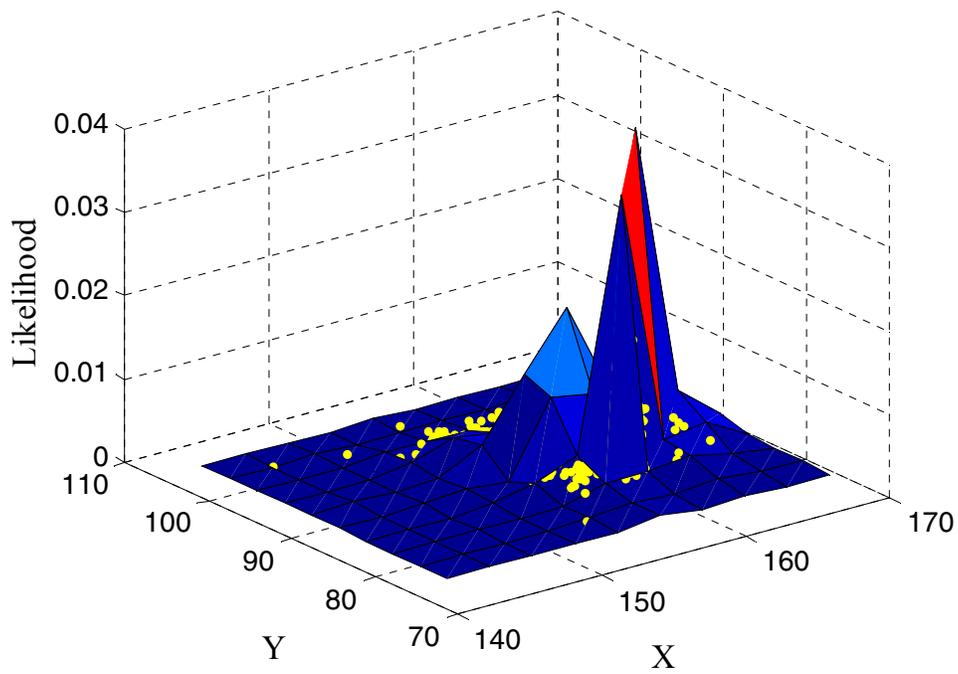

Fig.16 Li-PDF surf of fulcrums for the last frame (the yellow points represent the likelihood of particles that are calculated directly based on observations)

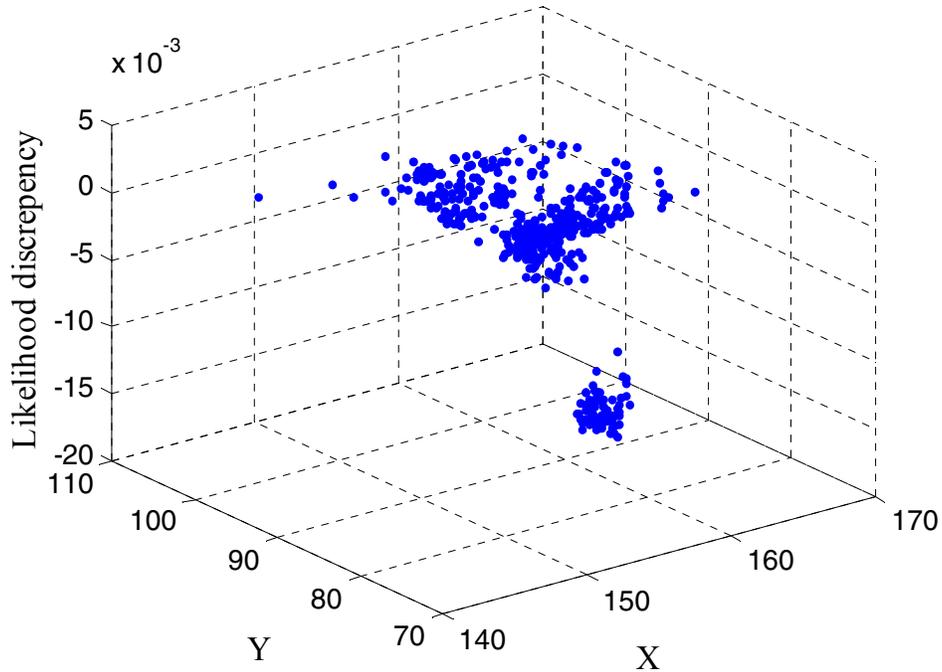

Fig.17 Discrepancy between the likelihood of particles that are interpolated by the Li-PDF approach and directly calculated based on observations

*4.4 Discussion*

Numerical fitting has been proven as a powerful tool for data prediction in many statistical applications as well as in our simulations. It is necessary to note that in most of known SSMs the observation function and the likelihood function actually appear to be easy for numerical fitting. The present Li-PDF approach can therefore be easily implemented. However, the simulations have exhibited its extra limitations for the likelihood calculation for which counter measures are needed.

First, direct numerical fitting may obtain negative likelihoods, which is not reasonable to update the particle weight that must be positive. To correct this, the negative likelihood can be set to zero.

Secondly, when a fixed number of fulcrums are used, more particles do not always lead to better estimation. To obtain the optimal approximation, the ratio of the number of fulcrums used to the number of particles should be set within a reasonable scope, for which we suggest the scope of [1/5, 1/2].

Thirdly, the Li-PDF approach is more suitable for relatively stable filtering in which the state transition is stable and the variance of the particle distribution is relatively small. Finding an ingenious solution to apply the numerical fitting approach to complicated state dynamics models and to situations when particles are distributed very widely in the state space still requires further research.

Finally, current Li-PDF applications use a fixed number of fulcrums and a constant form of fitting function which are simple but may not always be desirable since the complexity of the likelihood function often vary over time. Therefore, advanced/adaptive Li-PDF approaches that are able to adjust the number of fulcrums and the order of the fitting function would be valuable. This may be realized with the help of the online Goodness-of-fit test and is the subject of our future research. Further theoretical work will also include the convergence and stability analysis of the Li-PDF based particle filter.

## 5 Conclusions

This paper has proposed a novel particle filter that numerically fit the likelihood PDF (Li-PDF) in real time for speedy weight updating, which exhibits a significant mollification of the contradiction between the computational cost and the approximation accuracy in the particle filter. Using a relatively small number of fulcrums to infer the likelihood of particles, the Li-PDF approach alleviates the proportional dependence of the computational demand of PF on the number of particles used. It has been detailed how the numerical fitting method can be implemented on the observation function or directly on the likelihood function, namely implicit Li-PDF and explicit Li-PDF approaches. Exact simulation results have demonstrated that the processing speed of PF has been highly accelerated by the Li-PDF approach without losing the estimation accuracy. Meanwhile, an awareness of the limitations concerning the numerical fitting method for the Li-PDF calculation was also discussed with potential solutions provided.

In addition, the proposed numerical fitting method shows a potential ability for estimating the observation model. The future focus will be twofold: further development of advanced adaptive Li-PDF approaches that are capable of adjusting the number of fulcrums and the order of the fitting function for more challenging environments; and the employment of the numerical fitting tool within PF for system identification/parameter estimation.

## APPENDIXES

For engineering convenience, one may directly assume the Li-PDF function having linear dependence on the parameters as in (19). However, the assumption of the linearity dependence does not hold for most practical systems. This appendix gives the linearization error of converting a nonlinear fitting model to be one linear function. The analysis is based on the Taylor series expansion.

A Taylor series is a series expansion of a function about a point. A one-dimensional Taylor series of a real function $f(x)$ about a point $x=x_0$ is given by

$$f(x) = f(x_0) + f'(x_0)(x-x_0) + ... + \frac{f^{(n)}}{n!}(x-x_0)^n + R_n \quad (A.1)$$

where $R_n$ is a remainder term known as the Lagrange remainder, which is given by

$$R_n = \frac{f^{(n+1)}(x^*)}{(n+1)!}(x-x_0)^{n+1} \quad (A.2)$$

where $x^* \in [x_0, x]$ lies somewhere in the interval from $x_0$ to $x$.

Thus, if we use the engineering-friendly monomials function as in (19), there is at least a systematic error of $R_n$ occurring. This can be referred to as the system truncation error.

The expression $R_n$ in (A.2) indicates that the closer the prediction data $x$ is with $x_0$ (the smaller $(x-x_0)$ is), the more feasible it is to express the function in that interval in a lower order and with a smaller $R_n$. That is to say, the closer the particle is to the fulcrums (i.e. the smaller the piecewise interval), the more accurate and reliable the Li-PDF approach will be. This explains why the piecewise fitting is suggested in our approach to deal with the trade-off between smaller piecewise intervals and higher computational requirement.

## REFERENCES

[1] M. Šimandl, J. Královec, and T. Söderström. Anticipative grid design in point-mass approach to nonlinear state estimation, IEEE Trans. Automatic Control, vol. 47, no. 4, pp. 699-702, 2002.

[2] M. Šimandl, J. Královec, and T. Söderström. Advanced point-mass method for nonlinear state estimation, Automatica, vol. 42, pp. 1133–1145, 2006.

[3] T. Lefebvre, H. Bruyninckx, and J. de Schutter, Comment on a New Method for the Nonlinear Transformation of Means and Covariances in Filters and Estimators, IEEE Transactions on Automatic Control, vol. 47, no. 8, pp. 1406-1408, 2002.

[4] M. S. Arulampalam, S. Maskell, N. Gordon, and T. Clapp, A tutorial on particle filters for online nonlinear/non-Gaussian Bayesian tracking, IEEE Trans. Signal Process., vol. 50, no. 2, pp. 174–188, 2002.

[5] P. M. Djuric, J. H. Kotecha, J. Zhang, Y. Huang, T. Ghirmai, M. F. Bugallo, and J. Miguez, Particle filtering, IEEE signal processing magazine, vol. 20, no. 5, pp. 19–38, 2003.

[6] O. Cappe, and S. J. Godsill, and E. Moulines, An overview of existing methods and recent advances in sequential Monte Carlo, Proceedings of the IEEE, vol. 95, no. 5, pp. 899–924, 2007.

[7] A. Doucet, and A. M. Johansen, A tutorial on particle filtering and smoothing: Fifteen years later, in Handbook of Nonlinear Filtering, Ed. D. Crisan, and B. Rozovsky, Oxford: Oxford University Press, 2009.

[8] F. Gustafsson, F. Gunnarsson, N. Bergman, U. Forssell, J. Jansson, R. Karlsson, and P. Nordlund, Particle Filters for Positioning, Navigation and Tracking, IEEE Trans. Signal Process., vol. 50, no. 2, pp. 425-437, 2002.

[9] S.P. Won, W.W. Melek, and F. Golnaraghi, A Kalman/Particle Filter-Based Position and Orientation Estimation Method Using a Position Sensor/Inertial Measurement Unit Hybrid System, IEEE Trans. Ind. Electron., vol. 57, no.5, pp. 1787-1798, 2010.

[10] J. Lim, and D. Hong, "Cost Reference Particle Filtering Approach to High-Bandwidth Tilt Estimation," IEEE Trans. Ind. Electron., vol. 57, no. 11, pp. 3830-3839, 2010.

[11] H. Chen, and Y. Li, Enhanced particles with pseudo likelihoods for three-dimensional tracking, IEEE Trans. Ind. Electron., vol. 56, no. 8, pp. 2992-2997, 2009.

[12] W. R. Gilks, and C. Berzuini, Following a moving target—Monte Carlo inference for dynamic Bayesian models, J. R. Statist. Soc. B, vol. 63, pp. 127–146, 2001.

[13] A. Gning, B. Ristic, L. Mihaylova, F. Abdallah, Introduction to Box Particle Filtering, IEEE Signal Processing Magazine, vol. 30, no. 4, pp. 166 – 171, 2013.

[14] S. Thrun, W. Burgard, and D. Fox, "Probabilistic robotics," London: MIT Press, 2005. pp. 191-280.

[15] R. Havangi, H. D. Taghirad, M. A. Nekoui, and M.Teshnehlab, "A Square Root Unscented Fast SLAM With Improved Proposal Distribution and Resampling," *IEEE Trans. Ind. Electron.*, vol. 61, no. 5, pp. 2334-2345, 2014.

[16] J. Wang, Q. Gao, Y. Yu, H. Wang, and M.Jin, "Toward Robust Indoor Localization Based on Bayesian Filter Using Chirp-Spread-Spectrum Ranging," *IEEE Trans. Ind. Electron.*, vol. 59, no. 3, pp. 1622-1629, 2012.

[17] C. Chen, B. Zhang, G.Vachtsevanos, and M.Orchard, "Machine condition prediction based on adaptive neuro–fuzzy and high-order particle filtering," *IEEE Trans. Ind. Electron.*, vol. 58, no. 9, pp. 4353-4364, 2011.

[18] M. E. Orchard, P. Hevia-Koch, B. Zhang, and L. Tang, "Risk Measures for Particle-Filtering-Based State-of-Charge Prognosis in Lithium-Ion Batteries," *IEEE Trans. Ind. Electron.*, vol. 60, no. 11, pp. 5260-5269, 2013.

[19] C. Kwok, D. Fox, and M. Meilă, Real-time particle filters, Proceedings of the IEEE, vol. 92, no. 3, pp. 469-484, 2004.

[20] T. Li, S. Sun, and T. P. Sattar, Adapting sample size in particle filters through KLD-resampling, Electronics Letters, vol. 46, no. 12, pp. 740-742, 2013.

[21] T. Li, S. Sun, and T. P. Sattar, High-speed sigma-gating SMC-PHD filter, Signal Processing, vol. 93, no. 9, pp. 2586-2593, 2013.

[22] H. Pistori, V. Odakura, J. Bosco O. Monteiro, W. Gonçalves, A. Roel, J. Silva, and B. Machado, Mice and larvae tracking using a particle filter with an auto-adjustable observation model, Pattern Recognition Letters, vol. 31, no. 337-346, 2010.

[23] T. Li, M. Bolić, P. Djuric, Resampling methods for particle filtering, IEEE Signal Processing Magazine, 2014, to appear. Preprint is available online: https://sites.google.com/site/tianchengli85/publications/current-work/preprint

[24] G. Hendeby, R. Karlsson, and F. Gustafsson, Particle filtering: The need for speed, EURASIP Journal on Advances in Signal Processing, vol. 2010, no. 1-9, 2010.

[25] L. Mihaylova, A. Hegyi, A. Gning and R. Boel, Parallelized particle and Gaussian sum particle filters for large scale freeway traffic systems, IEEE Transactions on Intelligent Transportation Systems, vol. 13, no. 1, pp. 36 – 48, 2012.

[26] J. H. Kotecha and P. Djuric, Gaussian Particle filtering, IEEE Transaction on Signal Processing, vol.51, no.10, pp. 2592-2601, 2003.

[27] A. Doucet, S. Godsill and C. Andrieu. On sequential Monte Carlo sampling methods for Bayesian filtering, Statistics and Computing, vol. 10, pp. 197-208, 2000.

[28] T. Chen, T. B. Schön, H. Ohlsson, and L. Ljung, Decentralized particle filter with arbitrary state decomposition, IEEE Trans. Signal Process., vol. 59, no. 2, pp. 465-478, 2011.

[29] P. M. Djuric, T. Lu, and M. F. Bugallo, Multiple particle filtering, In: Proc. IEEE 32nd ICASSP, 2007, Pages. III-1181–III-1184.

[30] J. MacCormick, and M. Isard, Partitioned Sampling, Articulated Objects, and Interface-Quality Hand Tracking, In: Proceedings of the 6th European Conference on Computer Vision-Part II, 2000, pp. 3-19.

[31] D. Givon, P. Stinis, and J. Weare, Variance Reduction for Particle Filters of Systems With Time Scale Separation, IEEE Trans. Signal Process., vol. 57, no. 2, pp. 424-435, 2009.

[32] T. Li, S. Sun, Tariq P. Sattar and Juan M. Corchado, Fight sample degeneracy and impoverishment in particle filters: a review of intelligent approaches, Expert Systems With Applications, vol. 41, no. 8, pp. 3944-3954, 2014.

[33] D. Crisan, P. Del Moral, and T. Lyons, Discrete filtering using branching and interacting particle systems, Markov processes and related fields, vol. 5, no. 3, pp. 293–318, 1997.

[34] M. K. Pitt, and N. Shephard, Filtering via simulation: Auxiliary particle filters, J. Amer. Statist. Assoc., vol. 94, no. 446, pp. 590–591, 1999.

[35] C. Musso, N. Oudjane, and F. LeGland, Improving regularised particle filters, In Sequential Monte Carlo Methods in Practice, A. Doucet, J. F. G. de Freitas, and N. J. Gordon, Eds. New York: Springer-Verlag, 2001.

[36] C. Chang and R. Ansari, Kernel particle filter for visual tracking, IEEE Signal Processing Lett., vol. 12, no. 3, pp. 242 – 245, 2005.

[37] F. Campillo, V. Rossi, Convolution Particle Filter for Parameter Estimation in General State-Space Models. IEEE Transactions on Aerospace and Electronic Systems, vol. 45, no. 3, pp. 1063-1072, 2009.

[38] T. Yang, P.G. Mehta, S. P. Meyn, Feedback particle filters, IEEE Tran. Automatic Control, vol. 58, no. 10, pp. 2465-2480, 2013.

[39] B. Han, Y. Zhu, D. Comaniciu, and L. S. Davis, Visual tracking by continuous density propagation in sequential Bayesian filtering framework, IEEE Trans. PAMI, vol. 31, no. 5, pp. 919-930, 2009.

[40] S. C. Kramer, and H. W. Sorenson, Recursive Bayesian estimation using piece-wise constant approximations, Automatica, vol. 24, no. 6, pp. 789-801, 1988.

[41] F. Hayashi, Econometrics, Princeton University Press, 2000.

[42] M. Ades, and P.J. Van Leeuwen, An exploration of the Equivalent-Weights Particle Filter, Q. J. Royal Meteorol. Soc., vol. 139, no. 672, pp. 820-840, 2012.

[43] N. Vaswan, Particle Filtering for large-dimensional state space with multimodal observation likelihood, IEEE Trans. Signal Process., vol. 56, no.10, pp. 4583-4597, 2008.

[44] S. J. Julier, J. K. Ulhmann, and H. F. Durrant-Whyte, "A new method for nonlinear transformation of means and covariances in filters and estimators," IEEE Trans. Automat. Control, vol. 45, no. 3, pp. 472–482, 2000.